\documentclass[a4paper, 11pt]{article}
\usepackage{amsfonts,slashed}
\usepackage{float}
\usepackage{cite}
\usepackage{latexsym}
\usepackage{amsfonts}
\usepackage{amsmath}
\usepackage{epsfig}
\usepackage{latexsym,amssymb}
\usepackage{amsmath,amssymb,amsthm}
\usepackage{hyperref}
\usepackage{graphicx,tikz}
\usepackage{multirow}
\usepackage{textcomp}
\usepackage{xcolor}
\usepackage{mathtools}
\usepackage{stmaryrd}
\usepackage{enumerate}
\usepackage[shortlabels]{enumitem}

\usepackage{ytableau}

\usepackage{color}
\definecolor{red}{rgb}{1,0,0}
\definecolor{green}{rgb}{0,0.6,0}
\definecolor{blue}{rgb}{0,0,1}
\newcommand{\blue}[1]{\textcolor{blue}{#1}}
\newcommand{\red}[1]{\textcolor{red}{#1}}
\newcommand{\green}[1]{\textcolor{green}{#1}}

\usepackage{tikz}
\def\checkmark{\tikz\fill[scale=0.4](0,.35) -- (.25,0) -- (1,.7) -- (.25,.15) -- cycle;} 

\usepackage{ifpdf}
\ifx\pdfoutput\undefined
\pdffalse
\usepackage{cite}
\else
\pdfoutput=1
\pdftrue
\fi
\DeclareFontFamily{OMS}{rsfs}{\skewchar\font'60}
\DeclareFontShape{OMS}{rsfs}{m}{n}{<-5>rsfs5 <5-7>rsfs7 <7->rsfs10 }{}
\DeclareSymbolFont{rsfs}{OMS}{rsfs}{m}{n}
\DeclareSymbolFontAlphabet{\Scr}{rsfs}

\numberwithin{equation}{section}
\def\be{\begin{equation}}
	\def\ee{\end{equation}}
\def\ba{\begin{array}}
	\def\ea{\end{array}}

\newcommand{\bea}{\begin{eqnarray}}
	\newcommand{\eea}{\end{eqnarray}}

\newcommand{\cA}{\mathcal{A}}
\newcommand{\cB}{\mathcal{B}}

\newcommand{\calT}{\mathcal{T}}
\newcommand{\bbT}{\mathbb{T}}

\newcommand{\uA}{{\underline{A}}}
\newcommand{\uB}{{\underline{B}}}
\newcommand{\uC}{{\underline{C}}}
\newcommand{\uD}{{\underline{D}}}
\newcommand{\uE}{{\underline{E}}}
\newcommand{\uF}{{\underline{F}}}
\newcommand{\uG}{{\underline{G}}}

\newcommand{\SU}[1]{\mathrm{SU}( #1 )}
\newcommand{\SL}[1]{\mathrm{SL}( #1 )}

\newcommand{\SO}[1]{\mathrm{SO}( #1 )}

\newcommand{\U}[1]{\mathrm{U}(#1)}

\newcommand{\USp}[1]{\mathrm{USp}(#1)}

\newcommand{\En}[1]{E_{#1(#1)}}

\newcommand{\cV}{{\cal V}}

\newcommand{\mbf}[1]{\mathbf{#1}}

\newcommand{\gL}{\mathcal{L}}
\newcommand{\gM}{\mathcal{M}}

\newcommand{\cU}{{\cal U}}
\newcommand{\UI}{\left(U^{-1}\right)}
\newcommand{\flt}[1]{\underline{#1}}
\newcommand{\fl}[1]{#1}
\newcommand{\cY}{{\cal Y}}
\newcommand{\cN}{{\cal N}}
\newcommand{\vg}{\mathring{g}}

\newcommand{\cVI}{\left(\cV^{-1}\right)}
\newcommand{\cT}{{\cal T}}
\newcommand{\MGrav}{\Big(\mathbb{M}_{\rm spin-2}\Big)}
\newcommand{\MGravNoB}{\mathbb{M}_{\rm spin-2}}
\newcommand{\MGravZ}{\mathbb{M}_{\rm spin-2}}
\newcommand{\MVec}{\Big(\mathbb{M}_{\rm spin-1}\Big)}
\newcommand{\MVecZ}{\mathbb{M}_{\rm vec-tensor}}
\newcommand{\MScal}{\Big(\mathbb{M}_{\rm spin-0}\Big)}
\newcommand{\MScalZ}{\mathbb{M}_{\rm spin-0}}
\newcommand{\MTForm}{\Big(\mathbb{M}_{\rm tensor}\Big)}
\newcommand{\MTFormSq}{\Big(\mathbb{M}^2_{\rm tensor}\Big)}

\newcommand{\Cas}{{\cal C}}

\begin{document}
\begin{titlepage}
\vfill
\begin{flushright}
	HU-EP-23/60
\end{flushright}

\vfill
\begin{center}
	{\LARGE \bf Cubic and higher-order supergravity couplings\\[0.25em] for AdS vacua using Exceptional Field Theory
	}\\[1cm]
	
	{\large\bf Bastien Duboeuf\,$^{a}{\!}$
		\footnote{\tt bastien.duboeuf@ens-lyon.fr}, Emanuel Malek\,${}^{b}{\!}$
		\footnote{\tt emanuel.malek@physik.hu-berlin.de}, Henning Samtleben\,${}^{a,c}{\!}$
		\footnote{\tt henning.samtleben@ens-lyon.fr} \vskip .8cm}
	
	{\it ${}^a$ ENSL, CNRS, Laboratoire de physique, F-69342 Lyon, France}\\ \ \\
	{\it  $^{b}$ Institut f\"{u}r Physik, Humboldt-Universit\"{a}t zu Berlin,\\
		IRIS Geb\"{a}ude, Zum Gro{\ss}en Windkanal 2, 12489 Berlin, Germany}\\ \ \\
	{\it  $^{c}$ Institut Universitaire de France (IUF), France}\\ \ \\
	
\end{center}
\vfill

\begin{center}
	\textbf{Abstract}
	
\end{center}
\begin{quote}
We show how to use Exceptional Field Theory to efficiently compute $n$-point couplings of all Kaluza-Klein modes for vacua that can be uplifted from maximal gauged supergravities to 10/11 dimensions via a consistent truncation. Via the AdS/CFT correspondence, these couplings encode the $n$-point functions of holographic conformal fields theories. Our methods show that these $n$-point couplings are controlled by the $n$-point invariant of scalar harmonics of the maximally symmetric point of the truncation, allowing us to show that infinitely-many $n$-point couplings vanish for any vacua of the truncation, even though they may be allowed by the remnant symmetry group of the vacua. This gives new results even for the maximally supersymmetric AdS$_5 \times S^5$, AdS$_4 \times S^7$ and AdS$_7 \times S^4$ vacua of string and M-theory, where we prove old conjectures about the vanishing of $n$-point extremal and near-extremal couplings.

Focusing in particular on cubic couplings for vacua of 5-dimensional gauged supergravity, we derive explicit universal formulae encoding these couplings for any vacuum within a consistent truncation. We use this to compute known and new couplings involving spin-0, spin-1, spin-2 for the AdS$_5 \times S^5$ vacuum of IIB string theory.

\end{quote}
\vfill
\setcounter{footnote}{0}

\end{titlepage}

\tableofcontents

\newpage

\section{Introduction}

Compactifications play a pivotal role in string theory, where they are ubiquitous and form the cornerstone of various theoretical developments. One of the key features of compactifications are towers of Kaluza-Klein (KK) modes, which encode the geometry of the compactification from the perspective of the lower-dimensional theories. Gaining access to the spectrum of KK modes and their interactions can be very valuable, especially in the context of holography. The AdS/CFT correspondance states that operators in a Conformal Field Theory (CFT), are in one to one correspondance with fields living in the AdS bulk. In particular, the correspondance maps $n$-point functions of the CFT to couplings of fields in the gravitational theory. For example, the KK masses of supergravity fluctuations around a background are matched with 2-point functions of single trace operators in the dual CFT, while the cubic couplings of KK fluctuations encode the 3-point functions of the CFT. An intriguing point of this relation, is its a strong/weak nature. When one of the theories is strongly coupled, the theory on the other side of the correspondance becomes weakly coupled. In particular, for us, the weakly-coupled supergravity description of the AdS bulk gives us access to the (single-trace sector of the) strongly-coupled CFT.

Unfortunately, calculating the Kaluza-Klein spectrum, let alone their couplings, can be a daunting and often impossible task. Symmetries play a pivotal role in simplifying this process \cite{Salam:1981xd}. However, when a compactification possesses few symmetries, deciphering the mass spectrum was impossible until recently, when Exceptional Field Theory (ExFT) emerged as a crucial tool.

ExFT is a reformulation of 10/11-dimensional supergravity, unifying the metric and flux degrees of freedom and thus making manifest a hidden exceptional symmetry of the theories \cite{Berman:2010is,Berman:2011cg,Coimbra:2011ky,Berman:2012vc,Coimbra:2012af,Hohm:2013pua,Hohm:2013vpa}, which usually only appears after compactifying to lower dimensions. Structurally, ExFT rewrites 10-/11-dimensional supergravity in a way that mirrors the maximally supersymmetric gauged supergravities. For example, for the $\En{6}$ ExFT relevant to consistent truncations to 5 dimensions that we will mostly focus on here, the Lagrangian reads
\begin{equation}
	{\cal L}_{\rm ExFT6} = \sqrt{|g|}\, \Big( \widehat{R}
	+\frac{1}{24}\,g^{\mu\nu}{\cal D}_{\mu}{\cal M}^{MN}\,{\cal D}_{\nu}{\cal M}_{MN}
	-\frac{1}{4}\,{\cal M}_{MN}{\cal F}^{\mu\nu M}{\cal F}_{\mu\nu}{}^N
	+\sqrt{|g|}{}^{-1}{\cal L}_{\rm top}
	-V(g,{\cal M})\Big) \,,
	\label{eq:ExFTE6}
\end{equation}
where $g$ is a 5-dimensional metric on the external space, $\gM_{MN} \in \En{6}/\USp{8}$ is known as the generalised metric and encodes all the 5-dimensional scalar fields and ${\cal F}_{\mu\nu}{}^M$ is the field strength of the 5-dimensional vector fields, transforming in the $\mathbf{27}$ of $\En{6}$. These fields appear via their kinetic terms, which take the standard form of 5-dimensional supergravity, as well as a 5-dimensional potential $V(g,\gM)$ and a topological term ${\cal L}_{\rm top}$. In particular, there is a 5-dimensional Einstein-Hilbert term involving a Ricci scalar, $\widehat{R}$, for $g$, a scalar kinetic term involving the 5-dimensional gauge covariant derivative, ${\cal D}_\mu$, involving the $\mathbf{27}$ 5-dimensional vector fields and gauging generalised diffeomorphisms, i.e. diffeomorphisms and gauge transformations on the compactification, and finally a vector kinetic term involving ${\cal F}_{\mu\nu}{}^M$. Thus, the fields and the $\En{6}$ ExFT action \eqref{eq:ExFTE6} resemble the fields of maximal 5-dimensional supergravity, with the importance difference that all fields still carry arbitrary dependence on all 10- or 11-dimensional coordinates. Importantly, to rewrite 10-/11-dimensional supergravity in the form \eqref{eq:ExFTE6} that makes $\En{6}$ manifest, the 10-/11-dimensional fields need to be dualised and reorganised appropriately.

The use of different variables has often led to new insights in physics, for example elucidating new structures or enabling simpler computations. Similarly here, the reorganisation of fields to make exceptional symmetries manifest as in \eqref{eq:ExFTE6}, has proven powerful in computing KK spectra around any vacuum that belongs to a maximally supersymmetric consistent truncation of 10-/11-dimensional supergravity, as recently shown in \cite{Malek:2019eaz,Malek:2020yue}. Using these techniques offered by ExFT, the full KK spectra have been computed for a variety of vacua that were completely out of reach of traditional techniques. This, in particular, includes vacua with little or no (super-)symmetry, and has yielded interesting insights. For example, the recent work on KK spectra has revealed that there infinitely-many unprotected operators whose conformal dimensions are finite even at strong coupling \cite{Bobev:2020lsk,Cesaro:2021haf}, that certain conformal manifolds are compact, even though this is not apparent from studying the holographic duals within a consistent truncation \cite{Giambrone:2021zvp,Giambrone:2021wsm,Cesaro:2021tna,Cesaro:2022mbu}, that non-supersymmetric AdS vacua that are stable within a consistent truncation can nonetheless be perturbatively unstable due to higher KK modes that do not form part of the consistent truncation \cite{Malek:2020mlk}, and that there are several examples of isolated and continuous families of non-supersymmetric AdS vacua that are perturbatively stable within the full 10-dimensional supergravity \cite{Guarino:2020flh,Giambrone:2021wsm,Eloy:2021fhc,Eloy:2023zzh,Eloy:2023acy}.
 
Recently, these ExFT techniques have also been extended to vacua beyond maximally supersymmetric consistent truncations, which are instead deformations triggered by higher KK excitations that do not form part of the truncation \cite{Duboeuf:2022mam,Duboeuf:2023dmq}. For example, this means that ExFT tools can now be used to access KK spectra around general backgrounds, which are deformations of $\cN=8$ vacua, but which are not necessary described as a consistent truncation around the latter.

With this progress in computing the KK mass spectrum, a natural question is whether the powerful ExFT formalism can help in computing $n$-point couplings of the KK modes, where even less is known. For example, even for the maximally supersymmetric AdS$_5 \times S^5$ vacuum, only some cubic couplings have been computed \cite{Freedman:1998tz,Lee:1998bxa,Arutyunov:1999en,DHoker:1999jke,Arutyunov:1999fb} and matched to $\cN = 4$ SYM by brute force calculations. However, these results are only for specific fields and no results are known for the general $n$-point couplings even for this highly symmetric vacuum. Indeed, there are 20-year old conjectures \cite{DHoker:2000xhf,DHoker:2000pvz} about the vanishing of extremal and near-extremal $n$-point for the maximally supersymmetric AdS$_5 \times S^5$, AdS$_4 \times S^7$ and AdS$_7 \times S^4$ vacua of string and M-theory, whose proof previously appeared completely inaccessible from the supergravity side.

In this paper, we show how to extend the ExFT techniques to higher order couplings. We will show that the ExFT field basis \eqref{eq:ExFTE6} leads to a streamlined computation for $n$-point couplings, drastically simplifying the laborious computations arising using traditional supergravity techniques. These traditional approaches typically involve complicated fluctuation Ans\"{a}tze, gauge fixing, and field redefinitions that even invoke new higher-derivative terms and generally lack much structure. By contrast, in ExFT, we simply expand the two-derivative action \eqref{eq:ExFTE6} using a fluctuation Ansatz which can immediately be used for any vacuum of a consistent truncation of a maximally supersymmetric consistent truncation. Not only is the computation much simplified, ExFT also reveals a universal structure underlying the $n$-point couplings, which for any vacuum of a maximally supersymmetric consistent truncations are controlled by the same $n$-point invariant of scalar harmonics of the maximally symmetric point. This implies that infinitely many couplings vanish for any vacuum of the consistent truncation, even when allowed by group theory. For the maximally supersymmetric AdS$_5 \times S^5$, AdS$_4 \times S^7$ and AdS$_7 \times S^4$ vacua, this allows us to prove the conjectured vanishing of extremal and near-extremal $n$-point bulk couplings \cite{DHoker:2000pvz,DHoker:2000xhf}. Moreover, focusing on 3-point functions in E$_{6(6)}$ ExFT, we will obtain universal formulae that can be used to easily compute the cubic couplings all KK modes for any vacuum of a 5-dimensional maximally supersymmetric consistent truncation. 

Our paper is organized as follow. In section 2, we first briefly review E$_{6(6)}$ ExFT and recall the Kaluza-Klein mass formulae for vacua of 5-dimensional gauged supergravity. We then specify to the AdS$_5\times S^5$ vacuum, where the mass matrices nicely combine into Casimirs of the symmetry group. In section 3, we extend the ExFT analysis to $n$-point couplings, where we show that they are controlled by certain invariants of $n$ scalar harmonics of the maximally symmetric points of the truncations. We show how this causes infinitely many $n$-point couplings to vanish, despite being allowed by the symmetry groups of the vacua. In section 4, we specialise to cubic couplings and work out the explicit universal formulae which are valid for any vacuum that belongs to a 5-dimensional ${\cal N}=8$ consistent truncation for several fields of the E$_{6(6)}$ ExFT. Finally in section 5, we apply our results to the ${\cal N}=8$ AdS$_5 \times S^5$ vacuum of IIB string theory and prove the conjecture that extremal and near-extremal $n$-point couplings vanish. Moreover, we use our universal formulae to compute cubic couplings for AdS$_5 \times S^5$, comparing those that were already known with the literature and presenting several new results.

\section{ExFT mass matrices}
We review how to compute the Kaluza-Klein masses of any vacuum of a maximal gauged supergravity that arises from a consistent truncation of 10-/11-dimensional supergravity. We focus here on E$_{6(6)}$ ExFT, applicable to vacua of 5-dimensional maximal gauged supergravity, since this will be the principal focus of our paper. However, the structure for other dimensions is very similar and can be found in \cite{Malek:2020yue}.

\subsection{Review of $\En{6}$ ExFT}
The $\En{6}$ ExFT is a reformulation of 10-/11-dimensional supergravity in terms of the following set of fields: a 5-dimensional metric $g_{\mu\nu}$, a vector field $\cA_\mu{}^M$ in the $\mathbf{27}$ of $\En{6}$, a 2-form $\cB_{\mu\nu\,M}$ in the $\overline{\mathbf{27}}$ of $\En{6}$ and scalars $\gM_{MN}$ parameterising the coset space $\En{6}/\USp{8}$. The gauge structure of the $\En{6}$ ExFT is encoded via the generalised Lie derivative, an $\En{6}$ generalisation of the ordinary Lie derivative. This is given explicitly by
\begin{equation} \label{eq:GenLie}
	\gL_{V} W^M = V^N \partial_N W^M - 6\, \mathbb{P}^M{}_N{}^K{}_L\, W^N\, \partial_K V^L + \lambda\, W^M \partial_N V^N \,,
\end{equation}
where $\lambda$ is the weight of the generalised vector field $W^M$ and
\begin{equation} \label{eq:ProjAdj}
	\mathbb{P}^M{}_N{}^K{}_L = \frac1{18} \delta^M_N \delta^K_L + \frac16 \delta^M_L \delta^K_N - \frac53 d^{MKP} d_{NLP} \,,
\end{equation}
is the projector onto the adjoint of $\En{6}$, with $d^{MNP}$ the totally symmetric invariant tensor of $\En{6}$, normalised such that
\begin{equation}
	d_{MPQ} d^{NPQ} = \delta_M^N \,.
\end{equation}
Finally, the $\partial_M$ derivatives are formal objects, subject to the section condition
\begin{equation} \label{eq:sc}
	d^{MNP} \partial_M \otimes \partial_N = 0 \,,
\end{equation}
when acting on any product of fields of the ExFT. The section condition \eqref{eq:sc} allows dependence on only 5 or 6 coordinates, corresponding to type II or 11-dimensional supergravity. We refer the reader to \cite{Hohm:2013vpa} and \cite{Malek:2020yue} for the action and more details on the $\En{6}$ ExFT.

\subsection{Consistent Truncation Ansatz}
A consistent truncation from 10-/11-dimensional supergravity to 5-dimensional maximal gauged supergravity is possible whenever we have a globally-defined generalised frame which closes into an algebra under the generalised Lie derivative \cite{Hohm:2014qga,Lee:2014mla}. Manifolds where this is possible are known as generalised Leibniz parallelisable spaces \cite{Lee:2014mla}. Thus, we have a twist matrix
\begin{equation}
	U_M{}^A \in {\rm E}_{6(6)} \,,
\end{equation}
and a scalar density $\rho$ of weight $-\frac15$ such that the combination
\begin{equation}
	\cU_A{}^M = \rho^{-1} \UI_A{}^M
\end{equation}
satisfies
\begin{equation}
	\gL_{\cU_A} \cU_B{}^M = X_{AB}{}^C\, \cU_{C}{}^M \,,
\end{equation}
with $X_{AB}{}^C$ constant. $X_{AB}{}^C$ is, in general, known as the intrinsic torsion of the generalised parallelisation \cite{Cassani:2019vcl}, and when it is constant, it becomes the embedding tensor of the 5-dimensional maximal supergravity.

The consistent truncation Ansatz arises by expanding all fields of the $\En{6}$ ExFT in terms of $\rho$ and $U_M{}^A$ according to their weight under generalised diffeomorphisms and $\En{6}$ index structure. Thus, we have
\begin{equation} \label{eq:TruncationAnsatz}
	\begin{split}
		g_{\mu\nu}(x,y) &= \rho^{-2}(y)\, \mathring{g}_{\mu\nu}(x) \,, \\
		\cA_\mu{}^M(x,y) &= \rho^{-1}(y)\, \UI_{\fl{A}}{}^M(y)\, A_{\mu}{}^{\fl{A}}(x) \,, \\
		\cB_{\mu\nu\,M}(x,y) &= \rho^{-2}(y)\, U_M{}^{\fl{A}}(y)\, B_{\mu\nu\,\fl{A}}(x) \,, \\
		\gM_{MN}(x,y) &= U_M{}^{\fl{A}}(y)\, U_N{}^{\fl{B}}(y)\, M_{\fl{A}\fl{B}}(x) \,,
	\end{split}
\end{equation}
where $M_{\fl{A}\fl{B}}(x) = \cV_{\fl{A}}{}^{\flt{A}}\, \cV_{\fl{B}}{}^{\flt{B}}\, \delta_{\flt{A}\flt{B}}$ is the scalar matrix parameterising the scalar coset space $\frac{\En{6}}{\USp{8}}$ of the 5-dimensional maximal gauged supergravity.

\subsection{Kaluza-Klein Fluctuation Ansatz}
Since the twist matrix $U_M{}^A$ defines a globally well-defined generalised frame, we can use it to rewrite any tensor fluctuations in terms of scalar fluctuations. In turn, an arbitrary scalar fluctuation can be expanded in terms of a complete set of functions, which we denote by $\cY_\Sigma(y)$. Thus, we are led to the following Ansatz for Kaluza-Klein fluctuations around the origin of consistent truncation defined by \eqref{eq:TruncationAnsatz}
\begin{equation} \label{eq:KKAnsatz}
	\begin{split}
		g_{\mu\nu}(x,y) &= \rho^{-2}(y) \Big( \vg_{\mu\nu}(x) + \sum_\Sigma \cY_\Sigma(y)\, h_{\mu\nu}{}^{\Sigma}(x) \Big) \,, \\
		\cA_\mu{}^M(x,y) &= \rho^{-1}(y)\, \UI_\fl{A}{}^M(y) \sum_\Sigma \cY_\Sigma(y)\, A_\mu{}^{\fl{A}\,\Sigma}(x) \,, \\
		\cB_{\mu\nu\,M}(x,y) &= \rho^{-2}(y)\, U_M{}^\fl{A}(y) \sum_\Sigma \cY_\Sigma(y)\, B_{\mu\nu\,\fl{A}}{}^{\Sigma}(x) \,, \\
		\gM_{MN}(x,y) &= U_M{}^\fl{A}(y)\, U_N{}^\fl{B}(y) \Big( \delta_{\fl{A}\fl{B}} + \bbT_{\alpha A}{}^{B} \sum_\Sigma \cY_\Sigma(y)\, \phi^{\alpha\Sigma}(x) \Big) \,,
	\end{split}
\end{equation}
where $\bbT_{\alpha A}{}^{B}$ correspond to the non-compact generators of $\En{6}$, with the index $\alpha = 1, \ldots, 42$ raised and lowered using the non-compact part of the $\En{6}$ Cartan-Killing metric.

In order to compute the Kaluza-Klein masses around some other vacuum of the 5-dimensional maximal gauged supergravity, we simply replace
\begin{equation}
	U_M{}^{\fl{A}} \rightarrow U_{M}{}^{\flt{A}} = U_M{}^{\fl{A}}\, \cV_{\fl{A}}{}^{\flt{A}} \,,
\end{equation}
where $\cV_{\fl{A}}{}^{\flt{A}}$ is the scalar coset representative corresponding to the vacuum we are interested in. However, we can continue using the same complete set of function, $\cY_\Sigma$, as at the scalar origin. Typically, these can be easily computed, e.g. for $S^n$ truncations they correspond to the $\SO{n+1}$ spherical harmonics. Since any vacuum of the consistent truncation has the same topology, the harmonics of the scalar origin will define a complete basis for our fluctuations. What is not obvious at this stage is that using this basis of the scalar origin is also extremely efficient for any vacuum of the consistent truncation, as we will see in section \ref{s:LevelDiag}.

So far, all known maximal consistent truncations with compact gauge groups correspond to $S^n$ truncations, with the scalar origin being the round $S^n$. Therefore, in the following, we will refer to the scalar origin as the round $S^n$ (or even $S^5$, as we will largely focus on 5-dimensional truncations in this paper, although our results can straightforwardly be applied to maximally supersymmetric truncations to other dimensions on $S^n$), even though there could be maximally supersymmetric truncations with compact gaugings beyond sphere topologies.

\subsection{Mass matrices} \label{s:MassMatrix}
By plugging in the Kaluza-Klein fluctuation Ansatz \eqref{eq:KKAnsatz} into the linearised equations of motion of $\En{6}$ ExFT, we can easily compute the mass matrices for any vacuum of the 5-dimensional consistent truncation. These mass matrices will be expressed in terms of the embedding tensor dressed by the scalar coset representative as well as the action of the dressed generalised frame on the scalar harmonics
\begin{equation}
	X_{\flt{A}\flt{B}}{}^{\flt{C}} = \cVI_{\flt{A}}{}^{\fl{A}}\, \cVI_{\flt{B}}{}^{\fl{B}}\, \cV_{\fl{C}}{}^{\flt{C}}\, X_{\fl{A}\fl{B}}{}^{\fl{C}} \,, \qquad \gL_{\cU_{\flt{A}}} \cY_{\Sigma} = - \cT_{\flt{A}\,\Sigma}{}^{\Omega}\, \cY_{\Omega} \,.
\end{equation}
From the closure of the algebra of generalised Lie derivatives, it follows that the $\cT_{\flt{A}\,\Sigma}{}^{\Omega}$ matrices that capture the action of the dressed generalised frame on the scalar harmonics generate the algebra defined by the dressed embedding tensor
\begin{equation} \label{eq:TAlgebra}
	\left[ \cT_{\flt{A}},\, \cT_{\flt{B}} \right] = X_{\flt{A}\flt{B}}{}^{\flt{C}}\, \cT_{\flt{C}} \,.
\end{equation}

\subsubsection{Immediate level diagonalisation} \label{s:LevelDiag}
Before even computing the mass matrices explicitly, we can already observe a dramatic simplification in this approach compared to the standard one. With the ExFT fluctuation Ansatz \eqref{eq:KKAnsatz}, the mass matrices will be expressed in terms of the embedding tensor and the matrices $\cT_{\flt{A},\Sigma}{}^{\Omega}$. Crucially, the only object acting on harmonics are $\cT_{\fl{A}\,\Sigma}{}^{\Omega}$ matrices, which appear in their dressed form as
\begin{equation}
	\cT_{\flt{A}\,\Sigma}{}^{\Omega} = \cVI_{\flt{A}}{}^{\fl{A}} \cT_{\fl{A}\,\Sigma}{}^{\Omega} \,.
\end{equation}
The $\cT_{\fl{A}\,\Sigma}{}^{\Omega}$ form representations of the round $S^5$ isometries, and therefore do not mix different levels of scalar harmonics. Similarly, the dressed $\cT_{\flt{A}\,\Sigma}{}^{\Omega}$ matrices will also not mix different $S^5$ KK levels. This means that the KK levels of the round $S^5$ are preserved for any other vacuum of the consistent truncation. As a result, it becomes efficient to organise the KK spectrum of any vacuum of the consistent truncation in terms of the KK levels defined by the round $S^5$, something that is miraculous from the perspective of the remnant symmetry group of the vacuum.

In particular, we immediately see that in the basis defined by our fluctuation Ansatz \eqref{eq:KKAnsatz}, the mass matrices will remain diagonal level-by-level \emph{for any vacuum} of the consistent truncation. This fact is completely miraculous without knowledge of the consistent truncation and ExFT structure since, generically, the remnant symmetry group of a vacuum will allow a lot of mixing between the different KK levels. This fact drastically simplifies the computation of the KK spectrum compared to the traditional supergravity approach, and is just one of the reasons why the ExFT approach is so powerful when studying vacua with small remnant symmetry groups. However, even more intriguingly, the same mechanism will give rise to a powerful structure to cubic and higher couplings, as we will see in section \ref{s:CubicMixing}.

\subsubsection{Spin-2 mass matrix}
The spin-2 mass matrix is simply given by
\begin{equation} \label{eq:MassMatrixGravitonE6}
	\MGrav_{\Sigma}{}^{\Omega} = - \cT_{\flt{A}\,\Sigma}{}^{\Lambda}\, \cT_{\flt{A}\,\Lambda}{}^{\Omega} \,,
\end{equation}
where the repeated downstairs $\flt{A}$ index convention refers to summation by $\delta$, e.g. for two tensors in the $\overline{\mathbf{27}}$,
\begin{equation}
	V_{\flt{A}}\, W_{\flt{A}} = V_{\flt{A}}\, W_{\flt{B}}\, \delta^{\flt{A}\flt{B}} \,.
\end{equation}

\subsubsection{Vector \& tensor mass matrix}
 The vector mass matrix is given by
\begin{equation} \label{eq:MassMatrixVectorE6}
	\MVec{}^{\flt{A}\,\Sigma}{}_{\flt{B}\,\Omega} = \frac{1}{12}\, \Pi_{\flt{A}\,\alpha}{}^{\Sigma}{}_\Lambda\, \Pi^\alpha{}_{\flt{B}}{}^{\Lambda}{}_{\Omega} \,,
\end{equation}
with
\begin{equation}\label{eq:defPiMatrix}
	\Pi^\alpha{}_{\flt{A}}{}^{\Sigma}{}_\Omega = - 2 \left( X_{\flt{A}}{}^\alpha\, \delta^\Sigma_\Omega - 6\, \bbT^\alpha{}_{\flt{A}}{}^{\flt{B}}\, \cT_{\flt{B}}{}^{\Sigma}{}_{\Omega} \right) \,,
\end{equation}
and its adjoint
\begin{equation}
	\Pi_{\flt{A}\,\alpha}{}^\Sigma{}_\Omega = \Pi_{\alpha\,\flt{A}\,\Omega}{}^{\Sigma} = - 2 \left( X_{\flt{A}\alpha}\, \delta^\Sigma_\Omega + 6\, \bbT_{\alpha\flt{A}}{}^{\flt{B}}\, \cT_{\flt{B}}{}^{\Sigma}{}_{\Omega} \right) \,.
\end{equation}
Here, we defined the non-compact projection of the embedding tensor in its adjoint indices
\begin{equation}
	X_{\flt{A}}{}^\alpha = X_{\flt{A}\flt{B}}{}^{\flt{C}}\, \bbT^\alpha{}_{\flt{C}}{}^{\flt{B}} \,.
\end{equation}

On the other hand, the tensor mass matrix is
\begin{equation}
	\MTForm{}_{\flt{A}\,\Sigma}{}^{\flt{B}\,\Omega} = \frac{1}{\sqrt{10}} \left( - Z^{\flt{A}\flt{B}} \delta_\Sigma^\Omega + 10\, d^{\flt{A}\flt{B}\flt{C}}\, \cT_{\flt{C}\,\Sigma}{}^{\Omega} \right) \,,
\end{equation}
where we used the antisymmetric combination of the intrinsic torsion given by
\begin{equation}
	Z^{\flt{A}\flt{B}} = 2\, d^{\flt{C}\flt{D}\flt{A}}\, X_{\flt{C}\flt{D}}{}^{\flt{B}} = - Z^{\flt{B}\flt{A}} \,.
\end{equation}

These two expressions can be combined to find the following interesting structure arising.
\begin{equation} \label{eq:MassVectorTensor}
	\begin{split}
		\MVec^{\flt{A}\,\Sigma}{}_{\flt{B}\,\Omega} + \MTFormSq_{\flt{A}\,\Sigma}{}^{\flt{B}\,\Omega} &= \mathbb{M}_{0\,\flt{A}}{}^{\flt{B}}\, \delta_\Sigma^\Omega + 2\, \mathbb{N}{}_{\flt{C}\flt{A}}{}^{\flt{B}} \cT_{\flt{C}\,\Sigma}{}^{\Omega} + \delta_{\flt{A}\flt{B}} \MGrav_{\Sigma}{}^{\Omega} \\		
		& \quad - \frac53 \pi_{\flt{A}\,\Sigma}{}^{\Lambda} \pi_{\flt{B}\,\Lambda}{}^{\Omega} \,,
	\end{split}
\end{equation}
where
\begin{equation} \label{eq:NGenerator}
	\begin{split}
		\mathbb{M}_{0\,\flt{A}}{}^{\flt{B}} &= \frac13 X_{\flt{A}}{}^\alpha\, X_{\flt{B}\alpha} + \frac{1}{10} Z^{\flt{A}\flt{C}}\, Z^{\flt{B}\flt{C}} \,, \\
		\mathbb{N}_{\flt{C}\flt{A}}{}^{\flt{B}} & = \mathbb{N}_{\flt{C}}{}^{\hat{\alpha}}\, \bbT_{\hat{\alpha}\,\flt{A}}{}^{\flt{B}} = \left( X_{\flt{C}}{}^{\hat{\alpha}} + 3 X_{\flt{D}\flt{E}}{}^{\flt{C}}\, \bbT^{\hat{\alpha}}{}_{\flt{D}}{}^{\flt{E}} \right) \bbT_{\hat{\alpha}\,\flt{A}}{}^{\flt{B}} \,,
	\end{split}
\end{equation}
are the 5-dimensional gauged supergravity contribution and the tensor $\mathbb{N}_{\flt{C}}{}^{\hat{\alpha}}$ which mixes the level-0 and higher-level contributions. The $\bbT_{\hat{\alpha}\,\flt{A}}{}^{\flt{B}}$ are the compact generators with $\hat{\alpha}$ labelling the $\mathbf{36}$ of $\USp{8}$ and raised/lowered with the Cartan-Killing metric. Moreover, $\pi_{\flt{A}\,\Sigma}{}^{\Omega} = \cT_{\flt{A}\,\Sigma}{}^{\Omega}$ corresponds to the coupling between vectors and spin-2 and thus $\pi_{\flt{A}\,\Sigma}{}^{\Lambda} \pi_{\flt{B}\,\Lambda}{}^{\Omega}$ only acts on the spin-1 Goldstones eaten by the massive spin-2 states. We see that the $\cT^2$ terms act on physical modes exactly like the spin-2 mass operator:
\begin{equation}
	\MGravNoB = - \cT_{\flt{A}} \cT_{\flt{A}} \,.
\end{equation}

\subsubsection{Scalar mass matrix}
The scalar mass matrix is given by
\begin{equation} \label{eq:MassMatrixScalarE6}
	\begin{split}
		\MScal{}^{\alpha\,\Sigma}{}_{\beta\,\Omega} &= \Big[ X_{\uA\uE}{}^{\uF} X_{\uB\uF}{}^{\uE} \, \bbT^\alpha{}_{\beta\uA}{}^{\uB}  \\
		& \left. \quad + \frac15 \left( X_{\uA\uE}{}^{\uF} X_{\uB\uE}{}^{\uF} + X_{\uE\uA}{}^{\uF} X_{\uE\uB}{}^{\uF} + X_{\uE\uF}{}^{\uA} X_{\uE\uF}{}^{\uB}  \right) \bbT^\alpha{}_{\beta\uA}{}^{\uB} \right. \\
		& \quad + \frac25 \left( X_{\uA\uC}{}^{\uE} X_{\uB\uD}{}^{\uE} - X_{\uA\uE}{}^{\uC} X_{\uB\uE}{}^{\uD} - X_{\uE\uA}{}^{\uC} X_{\uE\uB}{}^{\uD} \right) \bbT^\alpha{}_{\uA}{}^{\uB}\, \bbT_{\beta\uC}{}^{\uD} \Big] \delta^\Sigma_\Omega \\
		&\quad + 2\, \Big( \bbT^\alpha{}_{\flt{A}}{}^{\flt{B}} X_{\flt{A}\beta} - \bbT_{\beta\flt{A}}{}^{\flt{B}} X_{\flt{A}}{}^{\alpha} \Big) \, \cT_{\flt{B}}{}^{\Sigma}{}_\Omega - 2\, \big[ \bbT^\alpha,\bbT_\beta\big]{}_{\flt{A}}{}^{\flt{B}} 
		\,X_{\flt{CB}}{}^{\flt{A}} \, \cT_{\flt{C}}{}^\Sigma{}_\Omega \\
		& \quad -\delta^\alpha_\beta\,\cT_{\flt{A}}{}^\Sigma{}_\Lambda \cT_{\flt{A}}{}^\Lambda{}_\Omega + 12\, \bbT^\alpha{}_{\beta\flt{A}}{}^{\flt{B}}\, \cT_{\flt{A}}{}^\Sigma{}_\Lambda \cT_{\flt{B}}{}^\Lambda{}_\Omega \,,
	\end{split}
\end{equation}
where we introduced the notation for products of generators
\begin{equation}
	\bbT_{\alpha\beta \flt{A}}{}^{\flt{B}} = \bbT_{\alpha \flt{A}}{}^{\flt{C}}\, \bbT_{\beta \flt{C}}{}^{\flt{B}} \,.
\end{equation}

Again, we can simplify this by adding terms which only act on Goldstone modes, i.e. if we add $\Pi^{\alpha}{}_{\flt{A}}{}^{\Lambda}{}_{\Sigma} \Pi^{\flt{A}}{}_{\beta}{}^{\Lambda}{}_{\Omega}$. We find
\begin{equation} \label{eq:MassMatrixScalarNice}
	\begin{split}
		\MScal{}^{\alpha\,\Sigma}{}_{\beta\,\Omega} + \frac1{12} \Pi^{\alpha}{}_{\flt{A}}{}^{\Lambda}{}_{\Sigma} \Pi^{\flt{A}}{}_{\beta}{}^{\Lambda}{}_{\Omega} &= \mathbb{M}_0{}^\alpha{}_\beta\, \delta^\Sigma{}_\Omega + 2\,\mathbb{N}_{\flt{A}}{}^{\alpha}{}_\beta \cT_{\flt{A}}{}^{\Sigma}{}_{\Omega} + \delta^\alpha{}_\beta \MGrav^\Sigma{}_\Omega \,.
	\end{split}
\end{equation}
Here
\begin{equation} \label{eq:MassMatrixScalarClean}
	\begin{split}
		\mathbb{M}_0{}^\alpha{}_\beta &= X_{\uA\uE}{}^{\uF} X_{\uB\uF}{}^{\uE} \, \bbT^\alpha{}_{\beta\uA}{}^{\uB} +  \frac12 X_{\uA\uE}{}^{\uF} X_{\uB\uE}{}^{\uF}\,´ \bbT^\alpha{}_{\beta\uA}{}^{\uB} \\
		& \quad + X_{\uA\uC}{}^{\uE} X_{\uB\uD}{}^{\uE} (\bbT^\alpha)_{\uA}{}^{\uB}\, (\bbT_\beta)_{\uC}{}^{\uD} + \frac13 X_{\flt{A}}{}^\alpha\, X_{\flt{A}\beta} \,, \\
		\mathbb{N}_{\flt{A}}{}^{\alpha}{}_\beta &= \mathbb{N}_{\flt{A}}{}^{\hat{\alpha}} \,  \bbT_{\hat{\alpha}}{}^{\alpha}{}_{\beta} \,,
	\end{split}
\end{equation}
where $\mathbb{N}_{\flt{A}}{}^{\hat{\alpha}}$ is the object from \eqref{eq:NGenerator} that controls the mixing between levels 0 and higher levels.

\subsection{AdS$_5 \times S^5$ spectrum}
Using \eqref{eq:MassMatrixGravitonE6}, \eqref{eq:MassVectorTensor} and \eqref{eq:MassMatrixScalarClean}, we can straightforwardly compute the AdS$_5 \times S^5$ spectrum. In order to evaluate the mass matrices, let us note the explicit form of the embedding tensor at this vacuum
\begin{equation}\label{eq:XonSO6}
	X_{\uA\uB}{}^{\uC}\rightarrow \left\{ \begin{array}{ll}
		X_{ab,cd}{}^{ef} = 2\sqrt{2}\delta_{[a}^{[e}\delta^{\phantom{[a]}}_{b][c}\delta_{d]}^{f]} \\
		X_{ab}{}^{c u}{}_{d v} = -\sqrt{2}\delta_{[a}^{c}\delta^{\phantom{a}}_{b]d}\delta_{v}^{u} 
	\end{array}\right.
	\,,
\end{equation}
where we decompose $\En{6} \rightarrow \SL{6} \times \SL{2}$, such that the $\mbf{27} \rightarrow \mathbf{\left(15,1\right)} \oplus \mathbf{\left(\overline{6},2\right)}$, and we use $u, v = 1, 2$ for the doublet. The action of the ${\cal T}$ matrix on the harmonics is given by
\begin{equation} \label{eq:TSO6}
	{\cal T}_{ab,c}{}^d = \sqrt{2} \delta^{\phantom{d}}_{c[a} \delta_{b]}^{d} \,,
\end{equation}
on fundamental harmonics ${\cal Y}^a$, and naturally extended on the higher harmonics, with ${\cal T}^{a u}{}_\Sigma{}^\Lambda = 0$.

We can now immediately compute the spin-2 mass spectrum. From \eqref{eq:TSO6}, we see that
\begin{equation}
	{\cal T}_{\flt{A}\,\Sigma}{}^{\Omega} = \sqrt{2}\, t_{\flt{A}\,\Sigma}{}^\Omega \,,
\end{equation}
where $t_{\flt{A}\,\Sigma}{}^{\Omega}$ are the canonically normalised $\SO{6}$ generators acting on the scalar harmonics. Thus,
\begin{equation}
	\MGravNoB = - \cT_{\flt{A}} \cT_{\flt{A}} = 2\, \Cas_{\SO{6}} = \ell(\ell+4) \,,
\end{equation}
where $\Cas_{\SO{6}}$ is the $\SO{6}$ Casimir and we expressed the spectrum in terms of the level, $\ell$, of the spin-2 fluctuation. On the other hand, the level-0 mass operators for vector/tensors, $\mathbb{M}_{0\,\flt{A}}{}^{\flt{B}}$, and for the scalars, $\mathbb{M}_0{}^\alpha{}_\beta$, can be evaluated explicitly and is, in both cases, given by
\begin{equation} \label{eq:SO6Level0}
	\mathbb{M}_{0} = 8 - 2\, \Cas_{\SO{6}} - 8\, \Cas_{\U{1}} \,,
\end{equation}
where $\Cas_{\U{1}}$ denotes the Casimir of the compact $\U{1} \subset \SL{2}_{\rm S}$ that remains of the IIB S-duality.

Finally, the tensor that controls the mixing between level-0 and higher-levels, $\mathbb{N}_{\flt{A}\flt{B}}{}^{\flt{C}}$, is given by
\begin{equation} \label{eq:NRelation}
	\mathbb{N}_{\flt{A}\flt{B}}{}^{\flt{C}} = - X_{\flt{A}\flt{B}}{}^{\flt{C}} \,.
\end{equation}
In fact, this relation holds for any embedding tensor that lives in the $\mathbf{36}$ of $\USp{8}$, such as for the AdS$_5 \times S^5$ vacuum. From the algebra of the ${\cal T}$-matrices \eqref{eq:TAlgebra} and using \eqref{eq:NRelation}, we see that the $\mathbb{N}$-tensor is given in terms of the canonically normalised $\SO{6}$ generators by
\begin{equation} \label{eq:N-tensorSO6}
	\mathbb{N}_{\flt{A}\flt{B}}{}^{\flt{C}} = \sqrt{2}\, t_{\flt{A}\flt{B}}{}^{\flt{C}} \,.
\end{equation}

Putting the level-0 result \eqref{eq:SO6Level0} together with the relation \eqref{eq:N-tensorSO6}, this now allows us to evaluate the vector-tensor \eqref{eq:MassVectorTensor} and scalar mass spectra \eqref{eq:MassMatrixScalarNice}. For this, we need to disinguish the action of $\SO{6}$ generators on level-0 fields and the scalar harmonics, which we denote by $t^0$ and $t^h$, respectively. Thus, we obtain for the mass matrix $\mathbb{M}$, corresponding to $\MVecZ$ or $\MScalZ$, the following
\begin{equation}
	\begin{split}
		\mathbb{M} &= \mathbb{M}_0 + 2\, \mathbb{N}_{\flt{A}} {\cal T}_{\flt{A}} - {\cal T}_{\flt{A}} {\cal T}_{\flt{A}} \\
		&= 8 + 2\, t_{\flt{A}}^0\, t_{\flt{A}}^0 - 8\, \Cas_{\U{1}} + 4 \, t_{\flt{A}}^0 \, t_{\flt{A}}^{h} - {\cal T}_{\flt{A}} {\cal T}_{\flt{A}} \\
		&= 8 + 2 \left( t_{\flt{A}}^0 + t_{\flt{A}}^h \right) \left( t_{\flt{A}}^0 + t_{\flt{A}}^h \right) - 8\, \Cas_{\U{1}} - 2\, {\cal T}_{\flt{A}} {\cal T}_{\flt{A}} \\
		&= 8 - 2\, \Cas_{\SO{6}} - 8\, \Cas_{\U{1}} + 2\, \ell\left(\ell+4\right) \,,
	\end{split}
\end{equation}
where in going to the second line we replaced $\Cas_{\SO{6}}$ from $\mathbb{M}_0$ by the combination $- t_{\flt{A}}^0\, t_{\flt{A}}^0$, in going to the third line, we completed the square, and, in going to the final line, we recognised $\left(t_{\flt{A}}^0 + t_{\flt{A}}^h\right)$ as the $\SO{6}$ generator acting on the tensor product of the level-0 fields and scalar harmonics, corresponding to our field basis.

Let us thus summarise our results for the AdS$_5 \times S^5$ spectrum. We have shown that the mass matrices reduce for a field in a representation $\left[n,p,q\right]_{\SO{6}} \otimes \left[j\right]_{\U{1}}$ at $S^5$ level $\ell$ to
\begin{equation} \label{eq:AdS5S5Spectrum}
	\begin{split}
		\MGravZ &= - 2\, \Cas_{\SO{6}} + 2\,\ell(\ell+4) \,, \\
		\MVecZ &= 8 - 2\, \Cas_{\SO{6}} - 8\, \Cas_{\U{1}} + 2\,\ell(\ell+4) \,, \\
		\MScalZ &= 8 - 2\, \Cas_{\SO{6}} - 8\, \Cas_{\U{1}} + 2\,\ell(\ell+4) \,,
	\end{split}
\end{equation}
where
\begin{equation}
	\begin{split}
		\Cas_{\SO{6}}[n,p,q] &= \frac18 \left(4n^2 + 3p^2 + 3q(4+q) + 2p(6+q) + 4n(4+p+q)\right) \,, \\
		\Cas_{\U{1}}[j] &= j^2 \,,
	\end{split}
\end{equation}
are the $\SO{6}$ and $\U{1}$ Casimirs and $\ell$ denotes the $S^5$ level of the representation. Note that our expression \eqref{eq:AdS5S5Spectrum} in terms of the $\SO{6}$ and $\U{1}$ Casimirs only appears once we write the fields as a tensor product of the 5-dimensional gauged supergravity fields with scalar harmonics. The compact expressions \eqref{eq:AdS5S5Spectrum} are already a signal that this new field basis we are using is very efficient in computing KK spectra. As we will see in \ref{s:S5Couplings}, this field basis also simplifies the computation and the results of $n$-point couplings.

\section{ExFT $n$-point couplings}

\subsection{Fluctuation Ansatz}
Cubic, and higher, couplings of the lower dimensional supergravity can be accessed using similar techniques as the ones used for KK spectroscopy, reviewed above. To obtain these $n$-point couplings, we need to extend fluctuation ansatz of \eqref{eq:KKAnsatz} by developing it to higher order. The key change comes from the scalar fluctuation Ansatz, which around any vacuum of the consistent truncation, can be written as an exponential
\begin{equation} \label{eq:FluctuationAnsatz}
	\begin{split}
		\gM_{MN}(x,y) &= U_M{}^{\flt{A}}(y)\, U_N{}^{\flt{B}}(y) \, \exp\left( \bbT_{\alpha \flt{A}}{}^{\flt{B}} \sum_\Sigma \cY_\Sigma(y) \phi^{\alpha\,\Sigma}(x) \right) \,.
	\end{split}
\end{equation}
For example, for cubic couplings, we need the expansion to quadratic order, given by
\begin{equation} \label{eq:QuadraticAnsatz}
	\begin{split}
		\gM_{MN}(x,y) &= U_M{}^{\flt{A}}(y)\, U_N{}^{\flt{B}}(y) \, \Big( \delta_{\flt{A}\flt{B}} + \bbT_{\alpha \flt{A}}{}^{\flt{B}} \sum_\Sigma \cY_\Sigma(y)\, \phi^{\alpha\,\Sigma}(x) \Big. \\
		&\quad \Big. + \frac12 \bbT_{\alpha\beta \flt{A}}{}^{\flt{B}} \sum_{\Sigma,\,\Omega} \cY_\Sigma(y)\, \cY_\Omega(y)\, \phi^{\alpha\,\Sigma}(x)\, \phi^{\beta\,\Omega}(x) \Big) \,,
	\end{split}
\end{equation}
where we again used the notation for products of generators
\begin{equation}
	\bbT_{\alpha\beta \flt{A}}{}^{\flt{B}} = \bbT_{\alpha \flt{A}}{}^{\flt{C}}\, \bbT_{\beta \flt{C}}{}^{\flt{B}} \,,
\end{equation}
that we will similarly use for higher products in the following. Plugging the Ansatz \eqref{eq:KKAnsatz}, \eqref{eq:FluctuationAnsatz} into the equation of motion or the Lagrangian then immediately gives access to the $n$-point couplings.

\subsection{Structure of $n$-point couplings} \label{s:CubicMixing}
In section \ref{s:LevelDiag}, we observed that the only action on KK levels in the mass matrices comes from the ${\cT}_{\flt{A}\,\Sigma}{}^{\Lambda}$ matrices. Since these do not mix different KK levels, we immediately concluded that the mass eigenstates do not mix different KK levels (defined with respect to the round $S^5$) either, even though the remnant symmetry group of the vacuum might allow plenty of mixing.

We can apply the same reasoning to $n$-point couplings using any ExFT, not just that based on $\En{6}$, to find a powerful structure which implies that many couplings vanish, even though they might be allowed by the symmetry group of the vacuum being studied. This is because also in the $n$-point couplings, the fluctuation Ansatz \eqref{eq:KKAnsatz}, \eqref{eq:FluctuationAnsatz} implies that the only derivatives of the scalar harmonics will come from the terms
\begin{equation}
	\gL_{\cU_{\flt{A}}} \cY_{\Sigma} = - \cT_{\flt{A}\,\Sigma}{}^{\Omega}\, \cY_{\Omega} \,,
\end{equation}
and thus will simply give rise to the dressed ${\cT}_{\flt{A}\,\Sigma}{}^{\Omega}$ matrices. All other derivatives will necessarily act on a twist matrix and give rise to a (dressed) embedding tensor
\begin{equation}
	X_{\flt{A}\flt{B}}{}^{\flt{C}} = \cVI_{\flt{A}}{}^{\fl{A}}\, \cVI_{\flt{B}}{}^{\fl{B}}\, \cV_{\fl{C}}{}^{\flt{C}}\, X_{\fl{A}\fl{B}}{}^{\fl{C}} \,.
\end{equation}
Therefore, we see that the only action on the scalar harmonics comes from the dressed ${\cT}_{\flt{A}\,\Sigma}{}^{\Omega}$ matrices, which do not mix different KK levels, with the level notion defined with respect to the $\SO{6}$ isometries of the round $S^5$.

We can, therefore, see that the $n$-point couplings will schematically appear in the action as
\begin{equation} \label{eq:cubicschematic}
	{\cal G}(\Phi^{{\cal A}_1\Sigma_1},\ldots, \Phi^{{\cal A}_n \Sigma_n}) \sim \int d\omega\, \Phi^{{\cal A}_1\Sigma_1}\, \Phi^{{\cal A}_2\Sigma_2}\, \ldots \Phi^{{\cal A}_n \Sigma_n} \left( \lambda_{{\cal A}_1{\cal A}_2 \ldots {\cal A}_n,\Sigma_1\Sigma_2\ldots\Sigma_n}{}^{\Delta_1\Delta_2 \ldots \Delta_n} \right) \cY_{\Delta_1} \cY_{\Delta_2} \ldots \cY_{\Delta_n} \,,
\end{equation}
where $\Phi^{{\cal A}_i\Sigma_i}$, $i = 1, \ldots, n$ denotes some generic KK excitation in a representation ${\cal A}_i$ of $\USp{8}$ and some 5-dimensional representation that we supressed and $d\omega$ represents the volume form of the compactification, while the $n$-point coupling is encoded in $\lambda_{{\cal A}_1{\cal A}_2 \ldots {\cal A}_n \,\Sigma_1\Sigma_2\ldots\Sigma_n}{}^{\Delta_1\Delta_2 \ldots \Delta_n}$, which is a quadratic expression of the dressed embedding tensor and the dressed ${\cT}_{\flt{A}\,\Sigma}{}^{\Omega}$ matrices. In \eqref{eq:cubicschematic}, the only internal dependence is now carried by the KK harmonics $\cY_{\Delta_1} \cY_{\Delta_2} \ldots \cY_{\Delta_n}$, whose integral gives the $n$-point invariant
\begin{equation}
	c_{\Delta_1\Delta_2\ldots \Delta_n} = \int d\omega\, \cY_{\Delta_1} \cY_{\Delta_2} \ldots \cY_{\Delta_n} \,.
	\label{eq:defc}
\end{equation}
Since we can use the same scalar harmonics for any vacuum of the consistent truncation, $c_{\Sigma_1\ldots\Sigma_n}$ is the same object for any vacuum. For example, for all vacua within the $\SO{6}$ gauged supergravity obtained by a consistent truncation on $S^5$, the cubic $c_{\Sigma\Delta\Gamma}$ is the unique cubic $\SO{6}$-invariant, up to normalisation.

Thus, the n-point couplings take the form
\begin{equation} \label{eq:cubicschematicsimple}
	{\cal G}(\Phi^{{\cal A}_1\Sigma_1},\ldots, \Phi^{{\cal A}_n \Sigma_n}) \sim \Phi^{{\cal A}_1\Sigma_1}\, \Phi^{{\cal A}_2\Sigma_2}\, \ldots \Phi^{{\cal A}_n\Sigma_n} c_{\Delta_1\Delta_2\ldots \Delta_n} \left( \lambda_{{\cal A}_1{\cal A}_2 \ldots {\cal A}_n,\Sigma_1\Sigma_2\ldots\Sigma_n}{}^{\Delta_1\Delta_2 \ldots \Delta_n} \right) \,.
\end{equation}
Crucially, because $\lambda_{{\cal A}_1{\cal A}_2 \ldots {\cal A}_n,\Sigma_1\Sigma_2\ldots\Sigma_n}{}^{\Delta_1\Delta_2 \ldots \Delta_n}$ does not mix different KK levels, i.e. the $\Sigma_i$ are the same KK levels as $\Delta_i$, $i = 1, \ldots, n$, this implies that $c_{\Delta_1\Delta_2 \ldots \Delta_n}$ will vanish when $c_{\Sigma_1\Sigma_2\ldots \Sigma_n}$ vanishes. We immediately conclude that if the scalar harmonics of the $n$ fields $\Phi^{{\cal A}_i\Sigma_i}$, $i = 1, \ldots n$, do not yield a non-vanishing $n$-point invariant $c_{\Sigma_1\Sigma_2\ldots \Sigma_n}$, then their $n$-point couplings will vanish. Importantly, this vanishing of $n$-point couplings is controlled by scalar harmonics of the round sphere but holds for \emph{all vacua} of the consistent truncation. This, therefore, implies that infinitely many $n$-point couplings vanish for vacua whose remnant symmetry group would have allowed such couplings. In fact, as we will show in section \ref{s:S5Conjectures}, even for the round spheres, e.g. the maximally supersymmetric AdS$_5 \times S^5$, AdS$_4 \times S^7$ and AdS$_7 \times S^4$ vacua, our results imply that more couplings vanish than would be expected from the isometries of the round sphere. Instead, our results show that extremal and non-extremal $n$-point couplings vanish, which was conjectured in \cite{DHoker:2000pvz,DHoker:2000xhf}. The vanishing of these $n$-point couplings is completely hidden without the use of ExFT.

We see that having a consistent truncation not only implies that all the $n$-point couplings vanish between any $n-1$ modes of the truncation and a mode that is not part of the truncation. In addition, the consistent truncation leaves a remnant for all higher KK levels, whose $n$-point couplings vanish if the KK levels, inherited from the round sphere, do not give a non-vanishing $n$-fold invariant. This result hold for all consistent truncations to maximal gauged supergravity, not just those in five dimensions.

\section{Cubic couplings from $\En{6}$ ExFT} \label{s:Cubic}
We are now ready to explore the cubic couplings in detail using $\En{6}$ ExFT and focusing on vacua that can be uplifted from 5-dimensional maximal gauged supergravity.

\subsection{Scalar cubic couplings}
The cubic couplings of three scalar fields are obtained from the potential, expanding the scalar matrix $\gM$ \eqref{eq:FluctuationAnsatz} in the potential. As we discussed in \ref{s:CubicMixing}, the couplings will be quadratic in the dressed embedding tensor $X_{\flt{A}\flt{B}}{}^{\flt{C}}$ and the dressed ${\cal T}_{\flt{A}\Sigma}{}^\Omega$ matrix. Thus, schematically, the cubic result is given by
\begin{equation} \label{eq:CubicSchematic}
{\cal G}(\phi^{\alpha\Sigma}, \phi^{\beta\Delta}, \phi^{\gamma\Gamma}) \propto \phi^{\alpha \Sigma} \phi^{\beta \Delta} \phi^{\gamma \Gamma} \left(XX_{\alpha \beta \gamma\Sigma \Delta \Gamma}+X\calT_{\alpha \beta \gamma\Sigma \Delta \Gamma}+\calT\calT_{\alpha \beta \gamma\Sigma \Delta \Gamma}\right) \,,
\end{equation}
where $XX$, $X{\cal T}$ and ${\cal T}{\cal T}$ refer to terms that are quadratic in $X$, linear in $X$ and ${\cal T}$ and quadratic in ${\cal T}$, respectively. The $XX$ part of the result is simply the coupling between the level 0 of the KK towers, and therefore can be obtained using 5-dimensional gauged supergravity without any ExFT analysis. However, for the couplings involving higher KK levels, we need the $X{\cal T}$ and ${\cal T}{\cal T}$ terms and thus a complete ExFT analysis.

While the naive cubic couplings come from plugging the fluctuation Ansatz \eqref{eq:FluctuationAnsatz} into the ExFT potential, our fluctuation Ansatz includes unphysical Goldstone modes. Since we are not interested in these unphysical couplings, we can change the couplings involving Goldstone fields in an effort to simplify the cubic coupling formulae. The Goldstones modes are encoded in the $\Pi$ matrix that was introduced earlier in \eqref{eq:defPiMatrix}
\begin{equation}
	\mathcal{D}_\mu \phi = \partial_{´\mu}\phi + A_{\mu} \phi + \Pi A_{\mu} \,,
\end{equation}
and describes how the Goldstone scalars couple to massive vector fields. Adding corrections from $\Pi$ matrix to the cubic couplings will not change the physical couplings, so that we can harmlessly simplify the expressions using $\Pi$ matrix corrections.

We choose to add terms involving the $\Pi$ matrices in such a way as to simplify the ${\cal T}{\cal T}$ terms as much as possible, analogously to what we did when simplifying the mass matrices in \ref{s:MassMatrix}. The resulting ${\cal T}{\cal T}$ term in \eqref{eq:CubicSchematic} is then given by
\begin{equation} \label{eq:3scalarsTT}
	\calT\calT_{\alpha \beta \gamma\Sigma \Delta \Gamma} = \calT_{\uB\,\Sigma}{}^{\Lambda} \calT_{\uA\,\Lambda}{}^{\Omega} c_{\Delta \Omega\Gamma}\left(6 \mathbb{T}_{\gamma \alpha \beta \uB}{}^{\uA} - \frac{1}{2} \mathbb{T}_{\gamma \uB}{}^{\uA}\kappa_{\alpha \beta}\right) \,,
\end{equation}
with the $X\calT$ part given by
\begin{equation}\label{eq:3scalarsXT}
	\begin{split}
		X\calT_{\alpha \beta \gamma\Sigma \Delta \Gamma} = \frac{1}{6}  \mathcal{T}_{\uB\,\Sigma}{}^{\Lambda} c_{\Gamma \Delta \Lambda}\Bigg[
		&X_{\uA \uC}{}^{\uD}\left(
		-6\, \mathbb{T}_{[\alpha\gamma] \uD}{}^{\uC} \mathbb{T}_{\beta \uB}{}^{\uA}  
		- 6\, \mathbb{T}_{\alpha \uD}{}^{\uC} \mathbb{T}_{\beta\gamma \uB}{}^{\uA} \right) \\
		&- X_{\uB \uA}{}^{\uC}\left(6\, \mathbb{T}_{\beta \alpha \gamma \uC}{}^{\uA} + \frac{1}{8} \kappa_{\gamma \beta}  \mathbb{T}_{\alpha \uC}{}^{\uA}\right)\Bigg] \,.
	\end{split}
\end{equation} 
Note that because \eqref{eq:3scalarsTT} and \eqref{eq:3scalarsXT} are contracted with three scalars in \eqref{eq:CubicSchematic}, the r.h.s. of \eqref{eq:3scalarsTT} and \eqref{eq:3scalarsXT} should be viewed as symmetrised in the exchange of the pairs ${\alpha\Sigma}$, ${\beta\Delta}$, and ${\gamma\Gamma}$.

This $XX$ part of the couplings, which of the three parts is the longest and up to the Goldstone shifts corresponds to the 5-dimensional gauged supergravity result, reads
\begin{equation} \label{eq:3scalarsXX}
\begin{split}
XX_{\alpha \beta \gamma\Sigma \Delta \Gamma} &= c_{\Sigma \Delta \Gamma} \Bigg[ X_{\uA \uB}{}^{\uC}\,X_{\uD \uC}{}^{\uB}\, \left(\frac16 \, \mathbb{T}_{\beta \gamma \alpha \uA}{}^{\uD}+\frac16 \kappa_{\gamma \alpha} \mathbb{T}_{\beta \uA}{}^{\uD} \right) \\ 
  &\quad + X_{\uA \uB}{}^{\uC} \, X_{\uD \uE}{}^{\uF}\, \left(
	\frac38\delta^{\uB \uE}\, \mathbb{T}_{\alpha \uA}{}^{\uD}\, \mathbb{T}_{\beta \gamma \uF}{}^{\uC}  
	+ \frac{37}{72}\delta^{\uA \uD}\, \mathbb{T}_{\alpha \uB}{}^{\uE}\, \mathbb{T}_{\beta \gamma \uF}{}^{\uC} \right. \\
	&\left. +\frac{19}{24}\,\delta_{\uC \uF}\, \mathbb{T}_{\beta \uB}{}^{\uE}\, \mathbb{T}_{\alpha \gamma \uA}{}^{\uD} 
 	+ \frac{97}{72}\, \delta^{\uA \uD}\, \mathbb{T}_{\alpha \uF}{}^{\uC}\, \mathbb{T}_{\beta \gamma \uE}{}^{\uB} - \frac{23}{18}\, \delta^{\uA \uD}\, \delta^{\uB \uE}\, \mathbb{T}_{\beta \gamma \alpha \uC}{}^{ \uF} \right) \Bigg] \,.
\end{split}
\end{equation}
Once again, because this term is multiplied by three scalars in \eqref{eq:CubicSchematic}, the r.h.s. of \eqref{eq:3scalarsXX} should be symmetrised in the exchange of ${\alpha\Sigma}$, ${\beta\Delta}$, and ${\gamma\Gamma}$.

\subsubsection*{Further simplification for embedding tensors in the $\mathbf{36}$ of $\USp{8}$}
We can significantly simplify the previous results by introducing additional assumptions. Recall that the embedding tensor of 5-dimensional maximal gauged supergravity transforms in the $\mathbf{351}$ representation of E$_{6(6)}$. When breaking E$_{6(6)}$ to its maximal compact subgroup $\USp{8}$, the $\mathbf{351}$ decomposes into $\mathbf{315} \oplus \mathbf{36}$. We are interested in exploring the implications of the embedding tensor being solely in the $\mathbf{36}$. This arises, for instance, for the maximally supersymmetric AdS$_5\times S^5$ vacuum.

Let's start by examining the $X\calT$ terms in \eqref{eq:3scalarsXT}. As we emphasised above, the multiplication of \eqref{eq:3scalarsXT} by three scalars in \eqref{eq:CubicSchematic} implies that the r.h.s. of \eqref{eq:3scalarsXT} should be symmetrised when exchanging the index pairs ${\alpha\Sigma}$, ${\beta\Delta}$, and ${\gamma\Gamma}$. However, the contraction of $\calT$ with a $c$-symbol in \eqref{eq:3scalarsXT} imposes a hook symmetry on the harmonic indices, since the $c$-symbol is invariant under the action of $\cal T$. As a result, the adjoint indices $\alpha$, $\beta$, $\gamma$ must also appear in the hook symmetric fashion. Therefore, stripping the ${\cal T}$ term from \eqref{eq:3scalarsXT}, the remaining tensor $\Xi_{\uB\alpha\beta\gamma}$ must exhibit a hook symmetry on the adjoint indices and, since it must be some $\USp{8}$-invariant combination of the dressed embedding tensor $X$, it must reside in the $\mathbf{36}$. Given these requirements, we find that only one term remains in all of \ref{eq:3scalarsXT}. By conducting a similar analysis on the $XX$-like terms in \eqref{eq:3scalarsXT}, we arrive at a single remaining term. After calculating the relative coefficients, we obtain the general formula for the cubic couplings of scalars for vacua where the embedding tensor is in the $\mathbf{36}$
\begin{equation}\label{eq:3scalars36}
\begin{split}
{\cal G}(\phi^{\alpha\Sigma}, \phi^{\beta\Delta}, \phi^{\gamma\Gamma}) \propto \phi^{\alpha \Sigma} \phi^{\beta \Delta} \phi^{\gamma \Gamma} \Bigg[& 
-\frac{1}{6} c_{\Sigma \Delta \Gamma}X_{\uA \uB}{}^{\uC}X_{\uD \uC}{}^{\uB}\, \mathbb{T}_{\beta \gamma \alpha \uA}{}^{\uD} \\
&- \mathcal{T}_{\uB\,\Sigma}{}^{\Lambda} c_{\Lambda \Delta \Gamma}X_{\uA \uC}{}^{\uD}\, \mathbb{T}_{[\alpha\gamma] \uD}{}^{\uC} \mathbb{T}_{\beta \uB}{}^{\uA} \\
&+\calT_{\uB\,\Sigma}{}^{\Lambda} \calT_{\uA\,\Lambda}{}^{\Omega} c_{\Omega \Delta \Gamma}\left(6\, \mathbb{T}_{\gamma \alpha \beta \uB}{}^{\uA} - \frac{1}{2}\, \mathbb{T}_{\gamma \uB}{}^{\uA}\kappa_{\alpha \beta}\right)\Bigg] \, .
\end{split}
\end{equation}

\subsection{Couplings involving spin-1}
In this section, we will give the cubic couplings involving at least one vector field.

We begin with the cubic couplings between one vector and two scalar fields. These come from the scalar kinetic term in the Lagrangian, which is given by
\begin{equation} \label{eq:KinScalar}
    \mathcal{L}_{\rm kin,\, scalar} = \frac1{24} \sqrt{-g} g^{\mu \nu} \mathcal{D}_{\mu} {\gM_{MN}}  \mathcal{D}_{\nu}{ \gM^{MN}}\,,
\end{equation}
where $D_\mu = \partial_\mu - \gL_{A_\mu}$ is the 5-dimensional derivative covariantised with respect to the ExFT generalised diffeomorphisms. Since we are interested in vector-scalar-scalar couplings, we will take $g^{\mu\nu} = \vg^{\mu\nu}$ to be the background metric of AdS$_5$ spacetime. By plugging in the fluctuation Ansatz \eqref{eq:KKAnsatz}, \eqref{eq:FluctuationAnsatz} into \eqref{eq:KinScalar}, we find
\begin{equation}
\begin{split}
	{\cal G}(A, \phi, \partial \phi) \propto  A_{\mu}{}^{\uA \Sigma}\Big[& 2\,c_{\Omega \Lambda\Sigma} X_{\uA \uB}{}^{\uC}\, \partial^{\mu}{\phi^{\alpha\Omega}}\, \phi^{\beta\Lambda}\, \bbT^{\hat{\gamma}}{}_{\alpha \beta}\, \bbT_{\hat{\gamma} \uC}{}^{\uB} + 2\, c_{\Delta \Lambda\Sigma}\, \phi^{\beta \Omega}\, \partial^{\mu}{\phi^{\alpha \Delta}} \,\calT_{\uA\,\Omega}{}^{\Lambda}\, \kappa_{\alpha \beta} \\ & - 12\, c_{\Lambda \Omega \Delta} \, \partial^{\mu}{\phi^{\alpha\Lambda}}\, \phi^{\beta\Delta}\, \calT_{\uB\,\Sigma}{}^{\Omega}\, \bbT^{\hat{\gamma}}{}_{\beta \alpha}\, \bbT_{\hat{\gamma} \uA}{}^{\uB}\Big] \,,
\end{split}
\label{eq:apdp}
\end{equation}
where $\bbT$ are the generators of $E_6$, with $\alpha, \beta, \dots$ labelling the non-compact and $\hat{\alpha}, \hat{\beta}, \hat{\gamma}$ labelling the compact indices of $\En{6}$, i.e. respectively the $\mathbf{42}$ and $\mathbf{36}$ of $\USp{8}$ and we suppressed the indices on $A$, $\phi$ and $\partial \phi$ on the l.h.s. for simplicity.

We now turn to couplings involving a vector field as well as 2-forms. These come from the field equations for the field strength $\mathcal{F}$, which is defined as
\begin{equation}
\mathcal{F}_{\mu \nu}{}^{M} = 2 \partial_{[\mu}\mathcal{A}_{\nu]}{}^{M} - 2 \mathcal{A}_{[\mu}{}^{K} \partial_{K} \mathcal{A}_{\nu]}{}^{M} + 10\, d_{P L K}\, d^{M N K} \mathcal{A}_{[\mu}{}^{P} \partial_N \mathcal{A}_{\nu]}{}^{L} + 10\, d^{M N K} \partial_{N}\mathcal{B}_{\mu \nu K} \,.
\end{equation}
The corresponding field equations are
\begin{equation} \label{eq:VecEoM}
    \mathcal{D}\star \mathcal{F}_{M} \propto d_{M N K} \, \mathcal{F}^{N}\wedge \mathcal{F}^{K} \,,
\end{equation}
where we have suppressed the 5-dimensional indices $\mu, \nu = 1, \ldots, 5$. Here $\star$ is the 5-dimensional Hodge dual, $\wedge$ is the 5-dimensional wedge operator and ${\cal D}$ is the ExFT covariantised 5-dimensional derivative operator, which in \eqref{eq:VecEoM} acts as an exterior derivative on ${\cal F}_M$ on the l.h.s. 

Plugging in the fluctuation Ansatz \eqref{eq:KKAnsatz}, we find
\begin{equation}
\begin{split}
	{\cal G}(A,B,B) & \propto A_{\mu}{}^{\uA\Sigma}\, B_{\nu \rho \uB \Delta}\, B_{\sigma \tau \uC \Gamma}\, \epsilon^{\mu \nu \rho \sigma \tau}\left(
d_{\uA \uD \uE}\, d^{\uC \uF \uE}\, d^{\underline{DBQ}} \calT_{\uF}{}^{ \Gamma}{}_{\Lambda} \calT_{\underline{Q}}{}^{\Delta}{}_{\Omega}\, c_{\Sigma}{}^{\Omega \Lambda} \right.\\ &
\left. -\frac{1}{10} d_{\uA \uD \uE}\, d^{\underline{CQE}}\, d^{\uG \underline{H} \uD} \calT_{\underline{Q}}{}^{\Gamma}{}_{\Lambda} X_{\uG \underline{H}}{}^{\uB}\, c_{\Sigma}{}^{\Delta \Lambda} \right. \\ &
\left. -\frac{1}{10} d_{\uA \uE \uD}\, d^{\uB \uF \uD}\, d^{\uE \uG \underline{H}} \calT_{\uF}{}^{\Delta}{}_{\Lambda} X_{\uG \underline{H}}{}^{\uC}\, c_{\Sigma}{}^{\Lambda \Gamma} \right. \\ &
\left. +\frac{1}{100} d_{\uA \uD \uE}\, d^{\underline{FQD}}\, d^{\uG \underline{H} \uE}
X_{\underline{FQ}}{}^{\uB} X_{\uG \underline{H}}{}^{\uC}\, c_{\Sigma}{}^{\Gamma \Delta} \right)\,.
\end{split}
\end{equation}

Finally we work out the couplings of three vectors. There are couplings of type $AA\partial A$, which only come from the kinetic term
\begin{equation}
	\mathcal{L}_{\rm kin,\, vec} = -\frac1{4} \sqrt{-g} \gM_{MN} {\cal F}^{\mu\nu\,M} {\cal F}_{\mu\nu}{}^N \,,
\end{equation}
and are given by
\begin{equation}
	{\cal G}(A,A,\partial A) \propto A^{\mu \uA\Sigma} A^{\nu \uB\Omega} \partial_{[\mu} A_{\nu]}{}^{\uC\Lambda} \Big(2\,c_{\Sigma \Lambda \Delta}\calT_{\uA\,\Omega}{}^{\Delta} \delta_{\uB \uC} + 10\,c_{\Sigma \Delta \Lambda} d_{\uA \uB \uD} d^{\uF \uC \uD}\calT_{\uF\,\Omega}{}^{\Delta}
	+c_{\Sigma \Omega \Lambda} X_{[\uA \uB]}{}^{\uC}\Big) \,.
	\label{eq:AAdA}
\end{equation}

On the other hand, the $A \partial A \partial A$ couplings only come from the topological term. Because both derivatives act along the external 5-dimensional space, these couplings have the very simple structure
\begin{equation}
	{\cal G}(A, \partial A, \partial A) \propto \epsilon^{\mu\nu\rho\sigma\lambda} A_{\mu}{}^{\flt{A}\Sigma}\, \partial_{\nu} A_{\rho}{}^{\flt{B}\Lambda}\, \partial_\sigma A_{\lambda}{}^{\flt{C}\Delta}\, d_{\flt{A}\flt{B}\flt{C}}\, c_{\Sigma\Lambda\Delta} \,.
\end{equation}

\subsection{Couplings between spin-2 and scalars}
Here we give the couplings between the fluctuations of the metric, which we will denote by $h$, and two scalars. From the structure of the indices, there is only one term that one can write for these couplings
\begin{equation} \label{eq:Spin2Scalar}
 {\cal G}(h, \partial \phi, \partial \phi) \propto \frac16 \kappa_{\alpha \beta}\, \partial_{\mu} \phi^{\alpha \Sigma}\, \partial_{\nu} \phi^{\beta \Delta}\, h^{\mu \nu \Lambda}\, c_{\Sigma \Delta \Lambda} \,.
\end{equation}
It can be checked that this is what is indeed obtained by expanding the scalar kinetic term from the Lagrangian \eqref{eq:KinScalar}.

\section{Example : AdS$_5 \times$S$^5$} \label{s:S5Couplings}

In this section, we use the previously introduced formalism in order to compute couplings on the background AdS$_5 \times$S$^5$. 
This background preserves maximal supersymmetry, i.e.\ states fall into supermultiplets of ${\rm SU}(2,2|4)$ and transform in representations
$[n,p,q]$ of the ${\rm SO}(6)$ R-symmetry group. Table~\ref{tab:BPS} recapitulates the structure of the $\frac12$-BPS supermultiplets ${\cal B}_{[n,0,0]}$ into 
which the supergravity spectrum decomposes.

\begin{table}[tb]
{\scriptsize
\begin{tabular}{c|l} 
     $\Delta$ &\\ \hline
     $n+2$ & $[n+2,00]{(0\,0)}$ \\[.5ex]
     $n+\frac52$ & $[n+1,10]{(0\,\frac12)}+
[n+1,01]{(\frac12\,0)}$ \\[.5ex]
     $n+3$ & $[n,02]{(0\,0)}+[n,20]{(0\,0)}
+[n+1,00]{(0\,1)}+[n+1,00]{(1\,0)}
+[n,11]{(\frac12\,\frac12)}$ \\[.5ex]
     $n+\frac72$ & $[n,10]{(0\,\frac12)}+[n-1,12]{(0\,\frac12)}
+[n,01]{(\frac12\,0)}+[n-1,21]{(\frac12\,0)}+
[n,01]{(\frac12\,1)}+[n,10]{(1\,\frac12)}$ \\[.5ex]
     $n+4$ & $2\!\cdot\![n,00]{(0\,0)}+[n-2,22]{(0\,0)}
+[n-1,02]{(0\,1)}+[n-1,20]{(1\,0)}+2\!\cdot\![n-1,11]{(\frac12\,\frac12)}+[n,00]{(1\,1)}$
\\[.5ex]
$n+\frac92$ &
$[n-1,10]{(0\,\frac12)}+[n-2,12]{(0\,\frac12)}+
[n-1,01]{(\frac12\,0)}+[n-2,21]{(\frac12\,0)}
+[n-1,01]{(\frac12\,1)}+[n-1,10]{(1\,\frac12)}$ \\[.5ex]
     $n+5$ & $[n-2,02]{(0\,0)}+[n-2,20]{(0\,0)}
+[n-1,00]{(0\,1)}+[n-1,00]{(1\,0)}
+[n-2,11]{(\frac12\,\frac12)}$ \\[.5ex]
     $n+\frac{11}2$ & $[n-2,10]{(0\,\frac12)}+
[n-2,01]{(\frac12\,0)}$ \\[.5ex]
     $n+6$ & $[n-2,00]{(0\,0)}$ \\
\hline
\end{tabular}
}
\caption{$\frac12$-BPS supermultiplets ${\cal B}_{[n,0,0]}$ of ${\rm SU}(2,2|4)$ in ${\rm SO}(6)\times{\rm SO}(4)$ notation $[n_1,n_2,n_3](j_1,j_2)$
with Dynkin labels $n_i$, and $(j_1,j_2)$ denoting the spins of ${\rm SO}(4)\sim{\rm SU}(2)\times {\rm SU}(2)$.}
\label{tab:BPS}
\end{table}

As has been discussed above, our fluctuation ansatz (\ref{eq:KKAnsatz}) introduces a different way of labelling the Kaluza-Klein states
by a couple of ${\rm SO}(6)$ indices, $\Phi^{\mathcal{A}\Sigma}$ , of which the first index refers to the field content of the ${\cal N}=8$ supergravity multiplet while the second index runs over the scalar harmonics on $S^5$. The latter are defined as polynomials in the fundamental harmonics $\mathcal{Y}^a$, $a=1, \dots, 6$, (satisfying $\mathcal{Y}^a\mathcal{Y}^{a}=1$) as
\begin{equation}\label{eq:HarmonicsDef0}
    \mathcal{Y}^{I} = \mathcal{Y}^{a_1 ... a_n} = \mathcal{Y}^{(\!(a_1} ... \mathcal{Y}^{a_n)\!)}
    \equiv \mathcal{Y}^{(a_1} ... \mathcal{Y}^{a_n)} - {\rm traces}
    \,,
\end{equation}
and transform in the symmetric vector representations $[n,0,0]$. We refer to appendix~\ref{app:harmonics} for a discussion of the properties of these harmonics and more explicit formulas.

All fields $\Phi^{\mathcal{A}\Sigma}$ with the second index in a given ${\rm SO}(6)$ representation $[n,0,0]$ combine into the $\frac12$-BPS supermultiplet ${\cal B}_{[n,0,0]}$. For example, for the scalar fluctuations, the index $\alpha$ on $\phi^{\alpha \Sigma}$ counts the 42 scalars of ${\cal N}=8$ supergravity. Under ${\rm SO}(6)$ these decompose according to 
\begin{equation}\label{eq:E6branching}
\mathbf{42} \rightarrow \mathbf{1_{+2}} \oplus \mathbf{1_{-2}} \oplus \mathbf{10_{+1}} \oplus \mathbf{\overline{10}_{-1}} \oplus \mathbf{20_{0}} \,,
\end{equation}
where subscripts refer to the different ${\rm SO}(2)$ charges. Accordingly, the fluctuations $\phi^{\alpha\Sigma}$ carry representations
\begin{align}
\mathbf{20} \otimes [n,0,0] &= 
[n\!+\!2,0,0] \oplus \textcolor{green}{[n,0,0]} \oplus [n\!-\!2,0,0]\oplus \textcolor{blue}{[n,1,1]}\oplus \textcolor{blue}{[n\!-\!2,1,1]}\oplus[n\!-\!2,2,2]
\,,\nonumber\\
\left(\mathbf{10}\oplus\mathbf{\overline{10}} \right) \otimes [n,0,0] &= 
[n,0,2] \oplus [n,2,0] \oplus \textcolor{blue}{2\cdot [n\!-\!1,1,1]}\oplus [n\!-\!2,2,0]\oplus [n\!-\!2,0,2]
\,,\nonumber\\
\left(\mathbf{1}\oplus\mathbf{1}\right) \otimes [n,0,0] &= 2\cdot [n,0,0]
\,,
\label{eq:42[n00]}
\end{align}
where ${\rm SO}(2)$ charges are easily restored. Comparing to Table~\ref{tab:BPS}, not all of these representations correspond to physical scalar fields in the supermultiplet ${\cal B}_{[n,0,0]}$, rather the representations given in \textcolor{blue}{blue} and \textcolor{green}{green} in (\ref{eq:42[n00]}) appear as Goldstone modes for the massive spin-1 and spin-2 fields, respectively.

\subsection{Near extremal scalar $n$-point couplings}\label{s:S5Conjectures}

Let us consider the $n$-point couplings of scalar fields transforming in fully symmetric traceless vector representations of type $[m,0,0]$. Explicitly, we label such representations by an index $I$
\begin{equation}
[m,0,0]:\quad
{\cal R}^I={\cal R}^{(\!(i_1 \dots i_m)\!)}\,,\quad\mbox{and define}\;\;
|I| \equiv m
\,,
\end{equation}
where $a_1, \dots a_m$ label the fundamental vector representation of ${\rm SO}(6)$, and $(\!( \dots )\!)$ denotes traceless symmetrisation.
We denote the corresponding fields in (\ref{eq:42[n00]}) as\footnote{For $s^I$ and $t^I$ this corresponds to the notation introduced in \cite{Lee:1998bxa,Arutyunov:1999en}. In contrast, the $\phi^I$ of \cite{Arutyunov:1999en} correspond to the (green) spin-2 Goldstone modes in (\ref{eq:42[n00]}), whereas the $\phi_\pm^I$ in (\ref{eq:stp}) denote physical scalars.}
\begin{equation}
s^{I}:\; [n\!+\!2,0,0] \,,\qquad
t^{I}:\; [n\!-\!2,0,0]\,,\qquad
\phi^{I}_\pm:\; 2\cdot [n,0,0]
\,.
\label{eq:stp}
\end{equation}
In particular, the $s^I$ are the chiral primaries of the supermultiplet, while the $\phi^{I}_\pm$ carry non-trivial ${\rm SO}(2)$ charge.

We have seen in section~\ref{s:CubicMixing} that the parametrisation of scalar fluctuations as $\phi^{\alpha\Sigma}$ together with the structure of the ExFT action allows to deduce strong constraints on the existence of possible couplings. Let us first illustrate this for the chiral primaries $s^I$. Consider an $n$-point coupling between this fields, that we shall denote as
\begin{equation}
{\cal G}(s^{I_1}, s^{I_2}, \dots , s^{I_n})
\,.
\label{eq:Gsn}
\end{equation}
${\rm SO}(6)$ representation theory immediately poses the condition
\begin{equation}
|I_i| \le \sum_{j\not=i} |I_j| \quad
\forall i
\,, 
\label{triangle0}
\end{equation}
necessary for a non-vanishing coupling (\ref{eq:Gsn}), more precisely for the appearance of a singlet in the tensor product $\bigotimes_j {\cal R}^{I_j}$. However, due to the fact that the fields $s^I$ appear as
\begin{equation}
s^I \in \phi^{ab,\Sigma}\,,\quad |I|=|\Sigma|+2
\,,
\label{eq:sphi}
\end{equation}
in the ExFT action, the general structure of couplings (\ref{eq:cubicschematicsimple}) shows that a non-vanishing coupling (\ref{eq:Gsn}) in fact requires
the stronger constraint
\begin{equation}
|\Sigma_i| \le \sum_{j\not=i} |\Sigma_j| \quad
\forall i \qquad\Longleftrightarrow\qquad
|I_i| +2\,(n-2) \le \sum_{j\not=i} |I_j| \quad
\forall i
\,.
\end{equation}
Put differently, we can conclude that 
\begin{equation}
 \Big( \sum_{j\not=i} |I_j|\Big) - |I_i| \;\le\; 2\,(n-3) 
\qquad\Longrightarrow\qquad
{\cal G}(s^{I_1}, s^{I_2}, \dots , s^{I_n})=0
\,,
\label{eq:conjecture}
\end{equation}
thus the vanishing of extremal and near-extremal couplings. This precisely corresponds to the conjecture 
first stated in~\cite{DHoker:2000xhf}, based on the lowest order explicit results~\cite{Lee:1998bxa,Arutyunov:1999en,Arutyunov:1999fb},  and found necessary to match the factored structure of the near-extremal correlation functions in weakly-coupled ${\cal N}=4$ SYM via AdS/CFT. The above reasoning gives a proof of this conjecture for arbitrary $n$-point couplings.

The conjecture (\ref{eq:conjecture}) has been reviewed in~\cite{DHoker:2000pvz} and been put into the context of consistent truncations. As discussed above, the truncation of the full Kaluza-Klein spectrum on AdS$_5\times S^5$ to the ${\cal N}=8$ supergravity multiplet ${\cal B}_{[2,0,0]}$ is consistent. Another way of stating this property is that the supergravity fields do not source the (truncated) higher Kaluza-Klein states, or equivalently, the absence of couplings linear in higher Kaluza-Klein fields, i.e.
\begin{equation}
I_1>2\;\mbox{and}\;|I_{i\ge2}|=2   \qquad\Longrightarrow\qquad
{\cal G}(s^{I_1}, s^{I_2}, \dots , s^{I_n})=0
\,,
\end{equation}
given that $s^I$, $|I|=2$, is the chiral primary of the supergravity multiplet. With hindsight, this is nothing but a very particular case of the general structure (\ref{eq:conjecture}). In turn, the ExFT structure of the action together with the above construction shows how the existence of a consistent truncation in fact implies the absence of numerous potential couplings, far beyond the lowest supergravity multiplet.

In~\cite{DHoker:2000pvz}, the conjecture on vanishing near-extremal couplings has further been generalised to the compactifications of eleven-dimensional supergravity on AdS$_4\times S^7$ and AdS$_7\times S^4$. In the ExFT framework, similar to the computations presented here, such couplings can be derived within E$_{7(7)}$ and ${\rm SL}(5)$ ExFT \cite{Hohm:2013uia,Musaev:2015ces}, respectively. While the technical details differ, the structure of the chiral primaries is still of the form (\ref{eq:sphi}), now with respect to the $R$-symmetry SO(8) and SO(5), respectively. The same general pattern thus applies and the conjecture (\ref{eq:conjecture}) is proven also in these cases for arbitrary $n$-point couplings.

In order to demonstrate the power of the framework, we may straightforwardly generalise the argument to other scalar fields, 
proving the vanishing of several other near extremal couplings. Let us consider couplings of the form
\begin{equation}
{\cal G}(s^{I_1},  \dots , s^{I_m},t^{J_1}, \dots , t^{J_n})
\,,
\end{equation}
between scalars of type $s^I$ and $t^J$, both appearing in fully symmetric traceless vector representations.
As above, ${\rm SO}(6)$ group theory requires that
\begin{align}
&|I_i| \le \sum_{j\not=i} |I_j|  + \sum_{\ell} |J_\ell| \quad
\forall i
\,,\nonumber\\ 
&|J_k| \le \sum_{j} |I_j|  + \sum_{\ell\not=k} |J_\ell| \quad
\forall k
\,, 
\label{eq:st}
\end{align}
in order to allow for a non-vanishing coupling (\ref{eq:st}).
Again, from the general structure of couplings (\ref{eq:cubicschematicsimple}), together with the embedding of fields as
\begin{equation}
s^I \in \phi^{ab,\Sigma}\,,\quad |I|=|\Sigma|+2
\,,
\qquad
t^J \in \phi^{ab,\Sigma}\,,\quad |J|=|\Sigma|-2
\,,
\label{eq:stphi}
\end{equation}
into the fluctuation ansatz, we may by reasoning similar to the above obtain the far stronger statements
\begin{align}
 \Big( \sum_{j\not=i} |I_j|  + \sum_{\ell} |J_\ell| \Big) - |I_i| \;\le\; 2\,(m-n-3) 
\quad\Longrightarrow\quad
&{\cal G}(s^{I_1},  \dots , s^{I_m},t^{J_1}, \dots , t^{J_n}) = 0 
\,,\nonumber\\ 
 \Big( \sum_{j} |I_j|  + \sum_{\ell\not=k} |J_\ell| \Big) - |J_k| \;\le\; 2\,(m-n+1) 
\quad\Longrightarrow\quad
&{\cal G}(s^{I_1},  \dots , s^{I_m},t^{J_1}, \dots , t^{J_n}) = 0 
\,,
\end{align}
which imply the vanishing of many near-extremal couplings whose presence would be compatible with 
${\rm SO}(6)$ symmetry (\ref{eq:st}).
It is straightforward to generalise this pattern to $n$-point couplings which further include the fields $\phi_\pm^I$ from (\ref{eq:stp}).
Similarly, one may derive conditions for vanishing couplings involving the fields in other ${\rm SO}(6)$
representations within the multiplet ${\cal B}_{[n,0,0]}$ of Table~\ref{tab:BPS}.

\subsection{Cubic scalar couplings}

Having shown that the general structure of our ansatz and the ExFT action impose the vanishing of numerous couplings, we will now use the explicit 
formulas (\ref{eq:3scalars36}) obtained for the cubic couplings in order to compute explicit expressions for some of the non-vanishing cubic couplings.
Before evaluating these equations, let us consider the expansion of the kinetic term for the scalar fields.
\begin{equation}
\frac1{24}\,{\cal M}^{KL} \partial_\mu {\cal M}_{LM} {\cal M}^{MN} \partial^\mu {\cal M}_{NK}
\;\longrightarrow\;
\frac1{24}\,\partial_\mu \phi^{\alpha\Sigma} \partial^\mu \phi^{\beta\Lambda}\,\delta_{\alpha\beta} \,
{\cal Y}^\Sigma {\cal Y}^\Lambda
+ {\cal O}(\phi^4)\,.
\label{eq:kinetic_exp}
\end{equation}
We note that in our basis of fluctuations, the expansion of this term does not give rise to cubic terms of type $\phi\partial_\mu\phi\partial^\mu\phi$. This is because in our parametrisation of fluctuations, the kinetic term respects USp(8) invariance in the first index of the $\phi^{\alpha\Sigma}$, and there is no USp(8) invariant cubic tensor $d^{\alpha\beta\gamma}$ that could define such a term.

Let us further note that for the fields $s^{I}$ from (\ref{eq:stp}), the expansion of the kinetic term (\ref{eq:kinetic_exp}) simply yields a normalisation constant proportional to $z(n)$ defined in (\ref{eq:intYY}), (\ref{eq:defz}), with $n=|I|-2$. Comparing this to the normalisation used in \cite{Lee:1998bxa,Arutyunov:1999en} for the same fields (c.f.\ equations (3.15), (3.23) in \cite{Lee:1998bxa}, with $k=|I|$)
\begin{equation}
{\cal A}_I \propto \frac{k(k-1)(k+2)}{k+1}\,z(k) \propto \frac{k^2(k-1)^2}{(k+1)^2}\,z(n)
\,,
\label{eq:normA}
\end{equation}
where we are ignoring overall constants that are independent of $k$ and where we have used the relation  
\begin{equation}
   z(n)  = z(k-2) = 4\,z(k)\,\frac{(k+1)(k+2)}{k(k-1)} 
   \,.
\end{equation}
From (\ref{eq:normA}) we see that our fields $s^I$ are related to the fields of \cite{Lee:1998bxa,Arutyunov:1999en} by rescaling
\begin{equation}\label{eq:phitos}
	s^I \longrightarrow \tilde{s}^I \equiv \frac{k+1}{k(k-1)}\, s^I
	\,.
\end{equation}
where for clarity we denote by $\tilde{s}^I$ for the fields of \cite{Lee:1998bxa,Arutyunov:1999en}.
Such rescaling factors will become relevant when comparing the results for the cubic couplings.

Let us now turn to evaluating the general formula (\ref{eq:3scalars36}) for AdS$_5\times S^5$. 
Recall that (\ref{eq:3scalars36}) is the reduction of the general formulae \eqref{eq:CubicSchematic}, \eqref{eq:3scalarsTT}, \eqref{eq:3scalarsXT}, \eqref{eq:3scalarsXX}, to the class of vacua whose associated embedding tensor lives in the ${\bf 36}$ of USp(8), which is the case for the round $S^5$.
While this formula reduces the contributions of general cubic scalar couplings to four terms, it turns out that
depending on the type of scalar fields not even all terms may contribute to the result.
The towers of scalar fields have been listed in (\ref{eq:42[n00]}), organised according to the representation of the first index $\alpha$ on $\phi^{\alpha,\Sigma}$.
In turn, for a given Kaluza-Klein fluctuation the representation of the associated $\alpha$ the can be inferred from its ${\rm SO}(2)$ charge.
Cubic couplings only exist among states whose ${\rm SO}(2)$ charges $q_i$ add up to zero, and we find that for the different possible combinations only the following terms from (\ref{eq:3scalars36}) give non-vanishing contributions
\begin{equation}
\label{Table:AllowedTermsS5}
\begin{tabular}{ c || c | c | c | c }
 $\langle q_1,q_2,q_3\rangle$  & $XX$ & $X{\cal T}$ & $({\cal T}{\cal T})_1$ & $({\cal T}{\cal T})_2$ \\ 
  \hline
 $\langle0,0,0\rangle$ & \checkmark & \checkmark &\checkmark&\checkmark \\
 \hline
 $\langle0,+1,-1\rangle$ & $-$ & \checkmark &\checkmark&\checkmark \\
  \hline
 $\langle0,+2,-2\rangle$ & $-$ & $-$ &$-$&\checkmark \\
  \hline
 $\langle\pm1,\pm1,\mp2\rangle$ & $-$ & $-$ & \checkmark & $-$  \\
\end{tabular}
\end{equation}
Here, $XX$, $X{\cal T}$, $({\cal T}{\cal T})_1$, and $({\cal T}{\cal T})_2$ denote the four terms in (\ref{eq:3scalars36}). For example, if the charges of scalars in a cubic coupling 
\begin{equation}
{\cal G}\left(\phi_1^{\alpha\Sigma},\phi_2^{\beta\Delta},\phi_3^{\gamma\Gamma}\right)\,,
\end{equation}  
are $0$, $+1$, and $-1$, the associated representations in $\alpha, \beta, \gamma$ are ${\bf 20}$, ${\bf 10}$ and ${\bf\overline{10}}$, respectively. The existence of a non-vanishing first term in $XX$ in (\ref{eq:3scalars36}) would require the existence of an ${\rm SO}(6)$ invariant tensor in the tensor product of these three representations, which does not exist. Similarly, one deduces the absence of several terms for other charges.

As a concrete example. let us compute the cubic coupling among three chiral primary fields
\begin{equation}
{\cal G}(s^{I_1},s^{I_2},s^{I_3})
\,,
\end{equation}
with $s^I$ from (\ref{eq:stp}). According to (\ref{Table:AllowedTermsS5}), all terms in (\ref{eq:3scalars36}) are present. The structure of (\ref{eq:3scalars36}) shows that the final result will be proportional to 
the structure constants defined in (\ref{eq:defc})
\begin{equation}
c_{\Sigma_1\Sigma_2\Sigma_3} =  \int_{S^5} d\omega\, {\cal Y}_{\Sigma_1}{\cal Y}_{\Sigma_2}{\cal Y}_{\Sigma_3}
 = a(n_1,n_2,n_3)\,{\cal C}_{\Sigma_1\Sigma_2\Sigma_3}
\,,
\quad
n_i\equiv  |\Sigma_i|=|I_i|-2\,,
\end{equation}
with the objects on the r.h.s.\ explicitly defined in (\ref{eq:defa123}), (\ref{eq:defc123}).
Plugging these explicit tensors into (\ref{eq:3scalars36}), we find the result
\begin{equation}\label{eq:CouplingsssonS5}
\begin{split}
{\cal G}(s^{I_1},s^{I_2},s^{I_3})
& = a(n_1,n_2,n_3)\Big(\frac{\sigma}{2} + 2\Big)\Big(\frac{\sigma}{2}+1\Big)\,{\cal C}^{\Sigma_1\Sigma_2\Sigma_3}\,
\delta_{ac}\delta_{be}\delta_{df} \phi^{(\!(ab, \Sigma_1)\!)} \phi^{(\!(cd,\Sigma_2 )\!)} \phi^{(\!(ef, \Sigma_3)\!)} \\
&= a(n_1,n_2,n_3)\Big(\frac{\sigma}{2} + 2\Big)\Big(\frac{\sigma}{2} +1\Big)\,{\cal C}^{I_1I_2I_3} \, s^{I_1} s^{I_2} s^{I_3}
\,,
\end{split}
\end{equation}
with $\sigma=n_1+n_2+n_3$, and the symbols ${\cal C}^{\Sigma_1\Sigma_2\Sigma_3}$, ${\cal C}^{I_1I_2I_3}$ 
defined in (\ref{eq:defc123}) for different representations, $[n,0,0]$ and $[n+2,0,0]$, respectively.

In order to compare this rather compact result to the explicit expressions obtained in \cite{Lee:1998bxa,Arutyunov:1999en}, we first note the relation
\begin{equation}\label{eq:aktoan}
   a(n_1,n_2,n_3) = a(k_1,k_2,k_3) \,\frac{\alpha_1 \,\alpha_2\, \alpha_3}{8\, k_1 k_2 k_3 (k_1-1)(k_2 -1)(k_3-1)}\,\frac{\tilde\sigma}{2}\Big(\frac{\tilde\sigma}{2} +1\Big)\Big(\frac{\tilde\sigma}{2} +2 \Big)
   \,,
\end{equation}
for $k_i=|I_i|=n_i+2$, with moreover $\tilde{\sigma}\equiv k_1+k_2+k_3=\sigma+6$, and $\alpha_i \equiv \frac12\tilde\sigma-k_i$.
Further taking into account the rescaling (\ref{eq:phitos}), the cubic coupling (\ref{eq:CouplingsssonS5}) takes the form
\begin{equation}
\label{eq:CouplingsssCompare}
{\cal G}(s^{I_1},s^{I_2},s^{I_3})
 =  \frac{\tilde\sigma\,\alpha_1 \,\alpha_2\, \alpha_3\,a(k_1,k_2,k_3)}{16\,  (k_1+1)(k_2 +1)(k_3+1)}
\left(\!\left(\tfrac{1}{2}\tilde\sigma\right)^2 -1\right)\left(\!\left(\tfrac{1}{2}\tilde\sigma\right)^2 -4\right)
{\cal C}^{I_1I_2I_3} \, \tilde{s}^{I_1} \tilde{s}^{I_2} \tilde{s}^{I_3}
\,,
\end{equation}
and this indeed is the result obtained in \cite{Lee:1998bxa,Arutyunov:1999en}.
The non-trivial zeros in this explicit expression (for any $\alpha_i=0$) precisely illustrate the theorem (\ref{eq:conjecture})
which in the ExFT formulation follows from purely structural arguments.


\subsection{Couplings involving spin-1 fields}

Let us work out a few other examples involving the spin-1 fields. Similar to (\ref{eq:E6branching}), (\ref{eq:42[n00]}), the spin-1 fields
are parametrised as $A_\mu{}^{A,\Sigma}$ with the USp(8) index $A$ decomposing as
\begin{equation}\label{eq:E6branchingV}
\mathbf{27} \rightarrow \mathbf{6_{+1}} \oplus \mathbf{15_{0}} \oplus \mathbf{{6}_{-1}} \,,
\end{equation}
under ${\rm SO}(6) \times \SO{2}$. Accordingly, the fluctuations $A_\mu{}^{A\Sigma}$ carry representations
\begin{align}
\mathbf{15} \otimes [n,0,0] &= 
[n,1,1] \oplus \textcolor{green}{[n,0,0]} \oplus \textcolor{blue}{[n\!-\!1,0,2]} \oplus \textcolor{blue}{ [n\!-\!1,2,0]}\oplus [n\!-\!2,1,1]
\,,\nonumber\\
\left(\mathbf{6}\oplus\mathbf{{6}} \right) \otimes [n,0,0] &= 
\textcolor{blue}{2\cdot [n\!+\!1,0,0]} \oplus\textcolor{blue}{2\cdot [n\!-\!1,0,0]} \oplus  2\cdot[n\!-\!1,1,1] 
\,,
\label{eq:27[n00]}
\end{align}
where similarly to (\ref{eq:42[n00]}) we have  denoted in \textcolor{blue}{blue} and \textcolor{green}{green} the Goldstone modes
for massive tensors and spin-2 modes, respectively. Following, \cite{Arutyunov:1999en} we denote the vector fields in the $[n,1,1]$ and the $[n\!-\!2,1,1]$, as $a_\mu{}^{\underline{I}}$ and $c_\mu{}^{\underline{I}}$, respectively. We use the index $\underline{I}$ to label the `hook' representations $[m,1,1]$ with $|\underline{I}|\equiv m+1$.
Explicitly, focusing on the $a_\mu{}^{\underline{I}}$, this corresponds to an embedding 
\begin{equation}
A_{\mu}{}^{ab,\Sigma} = \frac12\left(a_\mu{}^{a,(\!(b\Sigma)\!)}-a_\mu{}^{b,(\!(a\Sigma)\!)} \right)
\equiv a_\mu{}^{\underline{I}}
\,,\quad |\underline{I}|=|\Sigma|+1
\,.
\label{eq:embAa}
\end{equation}
Here, $A_\mu{}^{ab,\Sigma}$ is part of the general vector fluctuations in the basis of (2.9), with the index $A$ decomposed according to (5.29) and the $[m,1,1]$ fluctuation living in the $\mathbf{15}$ part, as can be seen from \eqref{eq:27[n00]}. $a_\mu{}^{a,a_1 \dots a_{m+1}}$ denotes the part of the fluctuation in the $[m,1,1]$, i.e.\ it is symmetric traceless in $(\!(a_1 \dots a_m)\!)$ with $a^{(a,a_1 \dots a_{m+1})}=0$.

We can now evaluate the above results for cubic vector couplings on the AdS$_5\times S^5$ background.  
Starting from (\ref{eq:AAdA}) and restricting to the vector fields (\ref{eq:embAa}),
we find that these couplings reduce to
\begin{align}
\label{eq:aadaS5}
{\cal G}(a^{\underline{I}_1},a^{\underline{I}_2},\partial a^{\underline{I}_3}) 
\propto\;& a(n_1,n_2,n_3) (\sigma + 6) 
\, \delta_{ae}\delta_{bc}\delta_{df} \,{\cal C}^{\Sigma_1\Sigma_2\Sigma_3}
\, a_{\mu}{}^{a,(\!(b \Sigma_1)\!)}a_{\nu}{}^{c,(\!(d \Sigma_2)\!)}\partial_{[\mu}a_{\nu]}{}^{e,(\!(f \Sigma_3)\!)}
\,,\nonumber\\
&n_i \equiv |\Sigma_i|=|I_i|-1\,,\quad \sigma = n_1+n_2+n_3
\,.
\end{align}
The index contraction on the r.h.s.\ of this equation defines an ${\rm SO}(6)$ invariant tensor ${\cal C}^{\underline{I}_1\underline{I}_2\underline{I}_3}$,
however unlike (\ref{eq:defc123}) there is not a unique invariant structure appearing in this tensor product, i.e.\ the particular
contraction in the coupling (\ref{eq:aadaS5})  by itself carries non-trivial information.

Turning now to vector-scalar couplings, let us evaluate the general coupling (\ref{eq:apdp}) for a vector field (\ref{eq:embAa}) with two chiral primaries $s^{I_1}$, $s^{I_2}$ from (\ref{eq:sphi}). After some computation we obtain the result in the form
\begin{align} 
\label{eq:CouplingAphiphiS5}
{\cal G}(\partial s^{{I}_1},s^{{I}_2},a^{\underline{I}_3}) 
\propto\;& a(n_1,n_2,n_3) \, (\sigma + 4) \left(\delta_{ea}\delta_{fb}\delta_{gc}-\delta_{eb}\delta_{fa}\delta_{gc}\right)
{\cal C}^{\Sigma_1\Sigma_2\Sigma_3}\partial^{\mu} \phi^{(\!(ag,\Sigma_1)\!)} \phi^{(\!(bc,\Sigma_2)\!)} \,A_{\mu}{}^{e,\!(f,\Sigma_3)\!)}
\nonumber\\[1.5ex]
\equiv\;& a(n_1,n_2,n_3) \,(\sigma + 4) \,
{\cal C}^{I_1I_2\underline{I}_3} \;\partial^{\mu} s^{I_1} s^{I_2} \, a_{\mu}{}^{\underline{I}_3}
\,,
\nonumber\\[1ex]
&n_i \equiv |\Sigma_i|\,,\quad \sigma = n_1+n_2+n_3
\,,
\end{align}
where $|\Sigma_i|=|I_i|-2$, for $i=1, 2$, and $|\Sigma_3|=|I_3|-1$.
Here, the tensor ${\cal C}^{I_1I_2\underline{I}_3}$ is defined by the above equation as the (up to normalisation)
unique invariant structure in this tensor product. It corresponds to what has been called $T_{123}$ in equation (8.72) of \cite{Arutyunov:1999en}.
In order to compare the full coupling (\ref{eq:CouplingAphiphiS5}) to the results of \cite{Arutyunov:1999en}, we need the relation
\begin{align}
\label{eq:aint}
a(n_1,n_2,n_3) = &\; \frac{(k_3+1)(\alpha_3-\tfrac12)(\tilde{\sigma}+3)(\tilde{\sigma}+1)}{k_1k_2k_3(k_1-1)(k_2-1)}t(k_1,k_2,k_3)
\,,\nonumber\\[1ex]
& \;k_1\equiv|I_1|=n_1+2\,,\quad k_2\equiv|I_2|=n_2+2\,,\quad k_3\equiv|\underline{I}_3|=n_3+1\,,\nonumber\\
&\;\tilde{\sigma}\equiv k_1+k_2+k_3=\sigma+5\,,\quad \alpha_3 \equiv \tfrac12\tilde\sigma-k_3\,,
\end{align}
to the $t(k_1,k_2,k_3)$ defined in equation (8.71) of \cite{Arutyunov:1999en}. Furthermore, in order to match the normalisation of vector fields, we note that our kinetic term
\begin{equation}
{\cal M}_{MN}\,{\cal F}_{\mu\nu}{}^M {\cal F}^{\mu\nu N} \longrightarrow
{\cal F}_{\mu\nu}{}^{A\Sigma} {\cal F}^{\mu\nu A\Lambda}\,{\cal Y}^\Sigma {\cal Y}^\Lambda 
\propto
\frac{n+2}{n+1}\, f_{\mu\nu}{}^{a,(\!(b\Sigma)\!)} f^{\mu\nu a,(\!(b\Lambda)\!)}\,{\cal Y}^\Sigma {\cal Y}^\Lambda 
\end{equation}
where the field strengths $f_{\mu\nu}$ are defined in analogy to (\ref{eq:embAa}). Comparing this to the normalisation of the kinetic term of \cite{Arutyunov:1999en} (whose fields we denote by $\tilde{a}_\mu{}^{\underline{I}}$, $\tilde{f}_{\mu\nu}{}^{\underline{I}}$),
we find
\begin{equation}
\frac{n+2}{n+1}\,z(n)\,{f}_{\mu\nu}{}^{\underline{I}} {f}^{\mu\nu}{}^{\underline{I}}
=
\frac{k+1}{k+2}\,z(k)\,\tilde{f}_{\mu\nu}{}^{\underline{I}} \tilde{f}^{\mu\nu}{}^{\underline{I}}
\,,
\qquad
k=n+1
\,,
\end{equation}
with $z$ from (\ref{eq:defz}). This yields the relation
\begin{equation}
\tilde{a}_\mu{}^{\underline{I}} = \frac{n+3}{n+1}\,a_\mu{}^{\underline{I}}
\,.
\end{equation}
Putting everything together, we can rewrite the result (\ref{eq:CouplingAphiphiS5}) as
\begin{align} 
\label{eq:CouplingAphiphiCompare}
{\cal G}(\partial s^{{I}_1},s^{{I}_2},a^{\underline{I}_3}) 
\propto\;& 
\frac{(k_3+1)(\alpha_3-\tfrac12)(\tilde{\sigma}+3)(\tilde{\sigma}+1)(\tilde\sigma -1)}{(k_1+1)(k_2+1)(k_3+2)}\,t(k_1,k_2,k_3)  \,
{\cal C}^{I_1I_2\underline{I}_3} \;\partial^{\mu} \tilde{s}^{I_1} \tilde{s}^{I_2} \, \tilde{a}_{\mu}{}^{\underline{I}_3}
\,.
\end{align}
Indeed, this is the form in which this coupling has been found in \cite{Arutyunov:1999en}.

\subsection{Couplings involving spin-2 fields}

As a final example, let us compute the coupling between a spin-2 field and two scalar fields,
which again we take to be the chiral primaries $s^I$. It is straightforward to evaluate (\ref{eq:Spin2Scalar}) for this case as
\begin{align}\label{eq:CouplingsshS5}
{\cal G}(\partial s^{I_1},\partial s^{I_2},h^{I_3}) 
 \propto &\,
  a(n_1,n_2,n_3)\,\partial_{\mu} \phi^{(\!(a b, \Sigma_1)\!)} \partial_{\nu} \phi^{(\!(ab, \Sigma_2)\!)}  h^{\mu \nu}{}^{\Sigma_3}\, 
 {\cal C}^{\Sigma_1\Sigma_2\Sigma_3}
 \nonumber\\[1.5ex]
 =&\,a(n_1,n_2,n_3)\,\partial_{\mu} s^{I_1}  \partial_{\nu} s^{I_2}\, h^{\mu \nu}{}^{I_3}\, 
 {\cal C}^{I_1I_2I_3}
 \,,
 \nonumber\\[1.5ex]
 &\,
 n_i=|\Sigma_i|
 \,,
\end{align}
where $|\Sigma_i|=|I_i|-2$ for $i=1, 2$, and $|\Sigma_3|=|I_3|$.
In order to compare this result to the literature, we apply the rescaling (\ref{eq:phitos}) as well as the 
relation
\begin{equation}
a(n_1,n_2,n_3) = a(k_1-2,k_2-2,k_3)= a(k_1,k_2,k_3)  \frac{(\tilde\sigma+4)(\tilde\sigma+2)}{k_1(k_1-1)k_2(k_2-1)}\,,
\end{equation}
for $k_1=n_1+2$, $k_2=n_2+2$ and $k_3=n_3$.
With this, the result (\ref{eq:CouplingsshS5}) turns into
\begin{align}
{\cal G}(\partial s^{I_1},\partial s^{I_2},h^{I_3}) 
=&\,
 \frac{(\tilde{\sigma}+4)(\tilde{\sigma}+2)(\alpha_3-1)\alpha_3}{(k_1+1)(k_2+1)}\,
 a(k_1,k_2,k_3) \,\partial_\nu \tilde{s}^{I_1} \partial_\mu \tilde{s}^{I_2} h^{\mu\nu I_3}\,{\cal C}^{I_1I_2I_3}
 \,,
 \nonumber\\[1.5ex]
 &\,
 k_i=|I_i|\,,\quad
 \tilde\sigma=k_1+k_2+k_3\,,\quad
 \alpha_3=\tfrac12\tilde\sigma-k_3
 \,,
\end{align}
which reproduces the result in the form found in \cite{Arutyunov:1999en}. 

\section{Conclusion}
In this paper, we showed how ExFT allows us to efficiently compute $n$-point couplings of KK modes around vacua of maximal gauged supergravities that arise as consistent truncations of 10/11-dimensional supergravity. These couplings are particularly interesting for AdS vacua, where they allow us to holographically compute correlators of single-trace operators of dual strongly-coupled CFTs. We showed that these $n$-point couplings have a remarkable structure, being controlled by an invariant $c_{\Sigma_1 \ldots \Sigma_n} = \int d´\omega\, {\cal Y}_{\Sigma_1} \ldots {\cal Y}_{\Sigma_n}$ that comes from integrating $n$ scalar harmonics and which is the same for any vacuum of the consistent truncation. This structure implies that infinitely many couplings vanish in supergravity, even though may be allowed by the symmetry group of the vacuum. This can be seen as a generalisation of the property of consistent truncation to higher KK levels. As we showed in \ref{s:S5Conjectures}, this structure even provides new results for the ${\cal N}=8$ AdS$_5 \times S^5$ vacuum, where this structure allows us to prove the previously conjectured vanishing of near-extremal couplings \cite{DHoker:2000xhf,DHoker:2000pvz}, as well as other vanishing $n$-point functions. For vacua that preserve fewer isometries, such as the ${\cal N}=2$ $\SU{2} \times \U{1}$-invariant AdS$_5$ vacuum of 5-d maximal $\SO{6}$ supergravity \cite{Khavaev:1998fb}, this structure is likely to lead to even more couplings vanishing.

In addition to investigating the structure of $n$-point couplings, we used ExFT to derive universal formulae that can be used to compute the cubic couplings of any vacuum of a consistent truncation to 5-dimensional maximal gauged supergravity. These couplings are controlled by specific quadratic combinations of the 5-dimensional embedding tensor and the action of Killing vectors on scalar harmonics, ${\cal T}$, and can be straightforwardly evaluated for any vacuum of the consistent truncation by appropriately dressing the embedding tensor and ${\cal T}$-matrices by the coset representative of the vacuum. We used this to compute cubic scalar, scalar-scalar-vector, scalar-scalar-spin-2 and cubic vector couplings, the first three of which match against the known literature while the last is a new coupling. Here another benefit of our approach becomes apparent: not only can the expressions be evaluated straightforwardly for any vacuum of the consistent truncation, but even for the round $S^5$ our field basis, consisting of tensor products of the 5-dimensional supergravity fields with the scalar harmonics, yields much more compact expressions for the couplings than the known results in the literature.

Our work opens up several avenues for further investigation. Since our method does not make any assumption about (super-)symmetries, our results can be used for vacua which preserve no isometries or are even non-supersymmetric. It would thus be interesting to investigate the consequence of the universal structure of $n$-point couplings and to explicitly compute the cubic couplings for less supersymmetric vacua, such as the ${\cal N}=2$ $\SU{2} \times \U{1}$ AdS$_5$ vacuum \cite{Khavaev:1998fb,Pilch:2000ej} dual to the Leigh-Strassler CFT \cite{Leigh:1995ep}. Together with the recently-computed KK spectrum \cite{Bobev:2020lsk}, this would effectively solve the single-trace sector of the Leigh-Strassler CFT at large $N$ and coupling.

Another direction would be to compute the universal cubic formulae using ExFT for other dimensions. For example, there are consistent truncations to several 4-dimensional gauged supergravities with interesting AdS vacua \cite{deWit:1986iy,Hohm:2014qga,Lee:2014mla,Guarino:2015jca,Inverso:2016eet,Ciceri:2016dmd,Cassani:2016ncu,Berman:2021ynm}, whose cubic couplings could be explored using $\En{7}$ ExFT. Yet another would be push the techniques presented here to work beyond vacua of consistent truncations to maximal supergravity, for example as was achieved for KK spectroscopy in \cite{Duboeuf:2022mam,Duboeuf:2023dmq}.

More generally, our work has uncovered a lot of structure in the $n$-point couplings for vacua that live in consistent truncations to maximal gauged supergravity. What does this surprising structure imply for dual CFTs? One would expect that the vanishing cubic couplings imply special OPE relations for the dual CFT. Can our results serves as useful input for the conformal bootstrap program or does the simplified basis and new structure arising from ExFT help with computing 4-point AdS correlators like in \cite{Rastelli:2017udc,Alday:2020dtb}? And, how much, if any, of this structure survives beyond the large-'t Hooft coupling and large-$N$ limits?

\section{Acknowledgements}
We are grateful to Nikolay Bobev and Hynek Paul for useful discussions and correspondence. EM is supported by the Deutsche Forschungsgemeinschaft (DFG, German Research Foundation) via the Emmy Noether program ``Exploring the landscape of string theory flux vacua using exceptional field theory'' (project number 426510644).

\appendix
\section{Harmonics}
\label{app:harmonics}

In this appendix we collect our conventions and some formulae for the harmonics on $S^5$. Starting from
the fundamental sphere harmonics ${\cal Y}^a$, $a=1, \dots, 6$, with ${\cal Y}^a{\cal Y}^a=1$, we define the
higher scalar harmonics by
\begin{equation}\label{eq:HarmonicsDef}
    \mathcal{Y}^{I} \equiv \mathcal{Y}^{a_1 ... a_n} \equiv \mathcal{Y}^{(\!(a_1} ... \mathcal{Y}^{a_n)\!)}
    =\mathcal{Y}^{(a_1} ... \mathcal{Y}^{a_n)} - {\rm traces}
    \,,\quad |I|\equiv n \,,
\end{equation}
with total weight 1 on the r.h.s.. 
These harmonics transform in the symmetric vector representation $[n,0,0]$ of ${\rm SO}(n)$.
We denote the integral over products of harmonics as
\begin{equation}
    \int_{S^5} d\omega\,\mathcal{Y}^{i_1} ... \mathcal{Y}^{i_n} \mathcal{Y}^{j_1} ... \mathcal{Y}^{j_n} = A_n \,m_n\,\delta^{(i_1i_2} \cdots \delta^{j_{n-1}j_n)}
    \,,
\end{equation}
where
\begin{equation}
    A_n = \frac{\pi^3}{2^{n-1} (n+2) !} \,,
\end{equation}
and
\begin{equation}
 m_n=   \frac{1}{n!} \binom{2n}{2} \cdots \binom{2}{2} = \frac{(2n)!}{2^n n!}
\,,
\end{equation}
counts the number of possibilities to distribute the indices over the $\delta$'s.
For the integral over a product of harmonics $\mathcal{Y}^{I} \mathcal{Y}^{J}$ with $|I|=|J|=n$, we then find 
\begin{equation}
\begin{split}
    \int_{S^5} d\omega\, \mathcal{Y}^{I} \mathcal{Y}^{J} & = 
    A_n \left(\delta^{(a_1}_{b_1} \cdots \delta^{a_n)}_{b_n}\right) n! - {\rm traces} \,,\\
    & = \frac{\pi^3}{2^{n-1}(n+1)(n+2)}\,\delta^{(\!(a_1}_{b_1} \cdots \delta^{a_n)\!)}_{b_n}
    \equiv z(n) \,\delta^{IJ}
    \,,
\end{split}
\label{eq:intYY}
\end{equation}
with
\begin{equation}
    z(n) =   \frac{\pi^3\, n!}{2^{n-1} (n+2)!}\,.
    \label{eq:defz}
\end{equation}
Similarly, the triple product of harmonics ${\cal Y}^{I_i}$ with $I_i=n_i$, $i=1,2,3$, is found to be
\begin{equation}\label{eq:Int3cY}
\begin{split}
    \int_{S^5} d\omega\, \mathcal{Y}^{I_{1}} \mathcal{Y}^{I_{2}} \mathcal{Y}^{I_{3}} & = 
    a(n_1,n_2,n_3) \, {\cal C}^{I_{1}I_{2}I_{3}}
    \,,
    \end{split}
\end{equation}    
with
\begin{equation}
    a(n_1,n_2,n_3) = \frac{n_1 ! n_2 ! n_3 !}{(\frac12 \sigma +2)! \, 2^{\frac12\sigma -1}}\frac{\pi^3}{\alpha_1! \alpha_2!\alpha_3! } \,,
    \label{eq:defa123}
\end{equation}
where $\sigma = n_1 + n_2 + n_3$, $\alpha_i = \frac{1}{2}\,\sigma-n_i$,
and ${\cal C}^{I_{1}I_{2}I_{3}}$ is the (up to normalisation) unique ${\rm SO}(6)$ invariant structure in the tensor product of the three representations, explicitly given by
\begin{equation}
{\cal C}^{\red{a_1 \dots ia_{n_1}},\blue{b_1 \dots b_{n_2}},\green{c_1 \dots c_{n_3}}} = 
\delta^{\red{a_1} \blue{b_1}} \dots \delta^{\red{a_{\alpha_3}} \blue{b_{\alpha_3}}}
\delta^{\red{a_{\alpha_3+1}} \green{c_1}} \dots \delta^{\red{a_{n_1}} \green{c_{\alpha_2}}}
\delta^{\blue{b_{\alpha_3+1}} \green{c_{\alpha_2+1}}} \dots \delta^{\blue{b_{n_2}} \green{c_{n_3}}}
\,,
\label{eq:defc123}
\end{equation}
where the indices in the same colour on the r.h.s.\ are understood to be projected onto the totally symmetric traceless part, such that the total weight is one. ${\rm SO}(6)$ group theory implies that a non-vanishing ${\cal C}^{I_{1}I_{2}I_{3}}$ requires the  triangle inequality
\begin{equation}
\alpha_i \ge 0
\,,
\end{equation}
which has been stated in equivalent form in (\ref{triangle0}).

\providecommand{\href}[2]{#2}\begingroup\raggedright\endgroup

\end{document}